\documentclass[reprint,aps,prb]{revtex4-2}

\usepackage{chemformula} 
\usepackage[T1]{fontenc} 
\usepackage{lmodern}
\usepackage{braket}
\usepackage{amsmath}
\usepackage{amssymb}
\usepackage{appendix}
\usepackage{hyperref}
\usepackage{epstopdf}
\usepackage{bm}
\usepackage{dcolumn}
\usepackage[version=4]{mhchem}

\usepackage{graphicx}
\usepackage[percent]{overpic} 
\usepackage{color}
\usepackage{transparent} 
\usepackage{ulem}

\usepackage[caption=false]{subfig}
\usepackage{circledtext}

\begin{document} 
	
	\title{Dynamics of antiskyrmion shrinking}
	
	\author{Frederik Austrup$^1$}
	\author{Wolfgang H{\"a}usler$^2$}
	\author{Michael Lau$^1$}
	\author{Michael Thorwart$^1$}
	\affiliation{$^1$I. Institut f{\"u}r Theoretische Physik, Universit{\"a}t Hamburg, Notkestra\ss{}e 9, 22607 Hamburg, Germany \\
		$^2$ Institut f{\"u}r Physik, Universit{\"a}t Augsburg, Universit{\"a}tsstra\ss{}e 1, 86135 Augsburg, Germany}
	
\begin{abstract}
Antiskyrmions are unstable in ferromagnetic systems with isotropic bulk or interfacial Dzyaloshinskii-Moriya interaction (DMI). We develop a continuum model for the shrinking dynamics of antiskyrmions in bulk DMI systems, using the Landau-Lifshitz-Gilbert equation for the time derivative of the magnetization field. Owing to the structure of their azimuthal angle, or helicity, elliptic antiskyrmions are energetically favored over circular ones. To capture this feature, we parametrize the magnetization field with a triangular radial profile and an elliptic in-plane shape. This ansatz yields four coupled dynamical equations governing time evolution of the semi-axes, helicities, and rotation angles. In the absence of the DMI, circular antiskyrmions shrink isotropically, exhibiting a crossover from exponential decay to square-root collapse. Initially elliptic antiskyrmions are driven towards circularity. For finite DMI, the semi-axes dynamics couples to the helicity and rotation, where the theory predicts a rotation angle following by half of the slope of the helicity evolution which is linear in time. Only at small semi-axes a cross-over to a logarithmic divergence occurs. The shrinking dynamics of the antiskyrmion size is found to be accompanied by quadrupole-like oscillations. Numerical simulations on the lattice support the predictions from the continuum model.
\end{abstract}

\maketitle
	
\section{Introduction}

Since their theoretical prediction \cite{BogdanovYablonskii,IvanovZhmudskii} and subsequent experimental discovery \cite{MuhlbauerBoni,YuTokura,HeinzeBlugel}, magnetic skyrmions and antiskyrmions have emerged as prime examples of topologically nontrivial spin textures in condensed matter systems. Their stability, particle-like character, and controllable dynamics allow manipulation by electric currents \cite{NagaosaTokura,IwasakiNagaosa}, magnetic fields \cite{ZhangHesjedal}, or thermal gradients \cite{QinChe,ChalusEskildsen} at low energy cost, making them promising candidates for both fundamental studies of magnetism and technological applications in spintronic devices based on individual skyrmions in thin-film materials \cite{KiselevRossler,TomaselloFinocchio,ZhangZhou,FinocchioKlaui,DupeHeinze,JiangHoffmann}.

Skyrmions can be stabilized by the antisymmetric Dzyaloshinskii-Moriya interaction (DMI) \cite{Moriya}, which arises in systems with broken inversion symmetry due to spin-orbit coupling. In chiral magnets with uniform DMI, skyrmions appear as stable configurations \cite{Everschor-SitteKlaui,Leonov}. Studies have shown that skyrmions can also be stabilized in the absence of DMI, for example in systems with uniaxial magnetic anisotropy \cite{LeonovMostovoy,Hou} or with higher-order exchange interactions \cite{Paul}. Antiskyrmions, in contrast, possess a twofold rotational symmetry. Along two orthogonal in-plane axes the magnetization shows Bloch- and anti-Bloch-type rotations, while along the axes rotated by $45^\circ$ it shows Néel- and anti-Néel-type  windings \cite{Koshibae}. This symmetry requires anisotropic DMI for stabilization \cite{CamosiVogel,Hoffmann,Nayak,CamosiRohart}, which can occur in materials with tetragonal $D_\text{2d}$ \cite{Nayak,PengTokura} or $S_4$ \cite{KarubeTaguchi,YasinYu} crystal structures. Such systems broaden the range of materials hosting topological spin textures and open pathways to novel magnetization dynamics and interactions. Recent studies have investigated current-driven antiskyrmion motion \cite{HeShen,Guang}. 

Also, pair-wise skyrmion-antiskyrmion creation from helical states under increasing magnetic fields \cite{Koshibae} and also via in-plane currents \cite{Stier} have been studied. In ferromagnets with isotropic bulk or interfacial DMI, skyrmions are energetically favored over the homogeneous state, whereas antiskyrmions remain unstable and eventually decay. Under pair creation, the skyrmion survives while the antiskyrmion collapses. In the absence of DMI and without additional stabilization mechanisms, both collapse together. This raises the question of how skyrmions and antiskyrmions shrink on their route to annihilation in these systems. For skyrmions, isotropic shrinking has been studied recently by us \cite{Austrup}. There, the dynamics is governed by two parameters only, the skyrmion radius and their helicity. The helicity rotates under an applied field, driving the system to alternations between Bloch- and Néel-type configurations. Since these configurations have different energies, helicity rotation is compensated by oscillations of the radius, leading to a  characteristic breathing dynamics. These have also been investigated in Ref.\  \cite{McKeeverEverschor-Sitte} for skyrmions and antiskyrmions, where the latter ones are studied in systems with anisotropic DMI, however, leaving the particular shrinking dynamics open.

In this work, we develop a continuum theory for the  dynamical shrinking of antiskyrmions in ferromagnets with isotropic bulk DMI, in which  skyrmions are stable but antiskyrmions are not. Compared to the skyrmion case, the decay of an antiskyrmion proves to be considerably more intricate. As described in Ref.\ \cite{Koshibae}, an antiskyrmion in a system with isotropic DMI tends to significantly distort from a circular shape due to direction-dependent energy costs or gains associated with different windings. Consequently, the polar profile of an antiskyrmion depends on the azimuthal angle. In our formulation, ellipticity lowers the energy by reducing the relative contributions of Bloch- and Néel-like regions within the helicity structure, making elliptic antiskyrmions favorable over circular ones. Consequently, the shrinking process involves additional degrees of freedom beyond the radius: the two semi-axes of the ellipse, the helicity, and the in-plane rotation angle now become time-dependent variables. 

We employ a triangular radial profile and an elliptic ansatz to investigate the energy landscape as a function of these parameters. Based on the Landau-Lifshitz-Gilbert equation \cite{LandauLifshits,Gilbert}, we derive four coupled dynamical equations which are highly non-linear. In the absence of DMI, the semi-axes decouple from the helicity and the rotation angle, leading to isotropic shrinking with a crossover from exponential decay at intermediate times to square root decay close to the collapse. In this regime, elliptic antiskyrmions evolve towards circularity. For finite DMI, however, the coupling between helicity, rotation, and the semi-axes drives quadrupole-like oscillations which overlay the decay. The helicity evolves linearly in time before diverging logarithmically near the collapse. In turn, the rotation angle follows at half the slope, modulated by step-like jumps synchronized with the semi-axes oscillations. All these features can be extracted from analyzing the coupled non-linear dynamical equations. The analytical predictions are supported by accurate numerical simulations on a grid. 

\section{Model}
\label{sec:Model}

We investigate the shrinking dynamics of an unstable antiskyrmion in a ferromagnetic host. The time evolution of the normalized magnetic moments $\bm{n}_i=\bm{n}_i(x,y,t)$ on a two-dimensional lattice, defined at lattice sites $i$, is governed by the Landau-Lifshitz-Gilbert (LLG) equation \cite{LandauLifshits,Gilbert}
\begin{equation}
\label{eq:LLGlattice}
\frac{\partial \bm{n}_i}{\partial t}=-\bm{n}_i \times \bm{B}_i^{\text{eff}} + \alpha \bm{n}_i \times \frac{\partial \bm{n}_i}{\partial t},
\end{equation}
where the first term on the right-hand side accounts for the precession of $\bm{n}_i$ around the local effective magnetic field $\bm{B}_i^{\text{eff}}$, which incorporates the Hamiltonian, and the second term describes the dissipative relaxation that eventually aligns $\bm{n}_i$ with $\bm{B}_i^{\text{eff}}$. For convenience, we choose the gyrocoupling constant $\gamma$ such that the natural unit of time is $t_0 = 1/J$ with $J$ being the exchange interaction strength (see below).  The damping strength is quantified by the phenomenological Gilbert parameter $\alpha$.

The Hamiltonian is characterized by the exchange interaction $J$, the DMI $D$, and an external magnetic field $B$ oriented along the $z$-axis. $B$ is expressed in dimensionless form and implicitly includes the saturation magnetization $M_s$, such that the LLG equation is written entirely in dimensionless units. These parameters enter the lattice model
\begin{equation}
\label{eq:Hamiltonian_general}
H = -\frac{J}{2} \sum_{\braket{i,j}} \bm{n}_i \bm{n}_j - D \sum_{\braket{i,j}} \bm{d}_{ij} \left[ \bm{n}_i \times \bm{n}_j \right] - B \sum_i n_i^z
\end{equation}
which defines the effective field via $\bm{B}_i^{\text{eff}}=-\partial H/\partial \bm{n}_i$. The summation $\braket{i,j}$ is restricted to nearest neighbors. The isotropic DMI is chosen to stabilize Bloch-type skyrmions and is specified through the vector $\bm{d}_{ij} = \frac{\bm{r}_j-\bm{r}_i}{|\bm{r}_j-\bm{r}_i|}$, where $\bm{r}_{j}=(x_j, y_j)$ denotes lattice coordinates in two dimensions \cite{Leonov}.

\section{Continuum Theory}
To analyze the shrinking dynamics of an antiskyrmion within an analytical framework, we adopt a continuum theory, following \cite{KronmullerFahnle}. In this approach, the LLG equation and the Hamiltonian in Eqs.\ \eqref{eq:LLGlattice} and \eqref{eq:Hamiltonian_general} are reformulated in terms of a classical field theory. The magnetization vector field is then parametrized to represent an antiskyrmion configuration.

\subsection{Energy functional and LLG equation of an antiskyrmion in the continuum limit}
We adopt the transition from the lattice to the continuum description as outlined in Ref.\  \cite{Austrup}. In this formulation, the lattice constant $a$ acts as the characteristic length scale that bridges the discrete lattice and the continuous field description and is absorbed into the system parameters, yielding $\tilde{D}=D/a$ and $\tilde{B}=B/a^2$. The continuum energy functional takes the form
\begin{equation}
\label{eq:EnergyFunctional}
\begin{split}
E=\int d^2r & \left[ \frac{J}{2} \left[ \left( \partial_x \bm{n} \right)^2 + \left( \partial_y \bm{n} \right)^2 \right] \right. \\
& \left. \vphantom{\frac{J}{2}} - \tilde{D} \bm{n} \left( \nabla \times \bm{n} \right) - \tilde{B} n_z \right].
\end{split}
\end{equation}
From this functional, the effective magnetic field in the continuum is obtained via the functional derivative of the energy density $\mathcal{E}$ with respect to $\bm{n}$ according to
\begin{equation}
\label{eq:Beff}
\bm{B}_{\text{eff}} = -\frac{\delta \mathcal{E}}{\delta \bm{n}} = J \Delta \bm{n} - 2 \tilde{D} \left(\nabla \times \bm{n} \right) + \tilde{B} \left( \begin{array}{c}
0 \\ 
0 \\ 
1
\end{array} \right),
\end{equation}
where $\mathcal{E}$ denotes the integrand of Eq.~\eqref{eq:EnergyFunctional}. Now, the LLG equation in the continuum limit reads
\begin{equation}
\label{eq:LLG}
\frac{\partial\bm{n}}{\partial t}  = -\frac{a^2}{1+\alpha^2} \left[ \bm{n} \times \bm{B}_\text{eff} - \alpha \bm{n} \times \left( \bm{n} \times \bm{B}_\text{eff} \right) \right].
\end{equation}

The magnetization vector is expressed using spherical coordinates as 
\begin{equation}
\label{eq:vectorfield}
\bm{n}=\left(\begin{array}{c}
\sin\Theta\cos\Phi  \\
\sin\Theta\sin\Phi  \\
\cos\Theta
\end{array} \right).
\end{equation}
Placing the origin of the coordinate system at the antiskyrmion center, the polar angle $\Theta(\rho,\varphi)$ depends on the radial distance $\rho$ and, additionally, in systems with isotropic bulk DMI on the direction $\varphi$, due to the two-fold rotational symmetry of the antiskyrmion. The azimuthal angle is assumed to be given as $\Phi(\varphi) = m\varphi + \varphi_0$, where $\varphi$ is the azimuthal coordinate and $\varphi_0$ denotes the helicity. An antiskyrmion corresponds to the case $m=-1$. The energy density $\mathcal{E} =\mathcal{E} (\rho,\Theta(\rho, \varphi),\Phi(\varphi))$ for such a skyrmion configuration reads
\begin{equation}
\label{eq:EnergyDensity}
\begin{split}
\mathcal{E} =& \frac{J}{2} \left[ (\partial_\rho\Theta)^2 + \frac{(\partial_\varphi \Theta)^2}{\rho^2} + \frac{\sin^2\Theta}{\rho^2} \right] \\
&+ \tilde{D} \left[ \left( \frac{\sin(2\Theta)}{2\rho} - (\partial_\rho\Theta) \right) \sin(2\varphi-\varphi_0) \right. \\ 
& \left. - \frac{ (\partial_\varphi\Theta)}{\rho}  \cos(2\varphi-\varphi_0) \right] \\
& + \tilde{B} - \tilde{B} \cos\Theta \, .
\end{split}
\end{equation}
Throughout this work, we measure $\tilde{D}$ and $\tilde{B}$ in units of $J$, therefore setting $J=1$. Also, the lattice constant is set to $a=1$ during all investigations.

\section{Description of an elliptic antiskyrmion}
Already from circularly symmetric skyrmions is known that the exact radial form of $\Theta(\rho)$ cannot be expressed in terms of analytical functions \cite{BogdanovHubert}. In practice, approximate expressions are often employed that reproduce the radial profile to high accuracy \cite{Wang}. A commonly used parametrization is
\begin{equation}
\label{eq:RadiusFit}
\Theta(\rho) = 2 \arctan\!\left[ \frac{\sinh(\rho_0 / w)}{\sinh(\rho / w)} \right],
\end{equation}
where $\rho_0$ denotes the skyrmion radius and $w$ characterizes the width of the domain wall-like shoulder in typical profiles. Although such functions provide quite accurate descriptions and will be used also in this work, they are not analytically tractable. Therefore, in the following we adopt a simpler, yet more manageable triangular profile \cite{BogdanovYablonskii}. Combined with a $\varphi$-dependent expression for the antiskyrmion radius $\rho_0(\varphi)$, this ansatz enables us to derive a set of coupled nonlinear ordinary differential equations that capture the shrinking dynamics of the antiskyrmion.

\subsection{Elliptic triangular approximation of the polar angle}
\label{sec:EllipticTriangularAnsatz}

\begin{figure}[!]
	\includegraphics[width=0.47\textwidth]{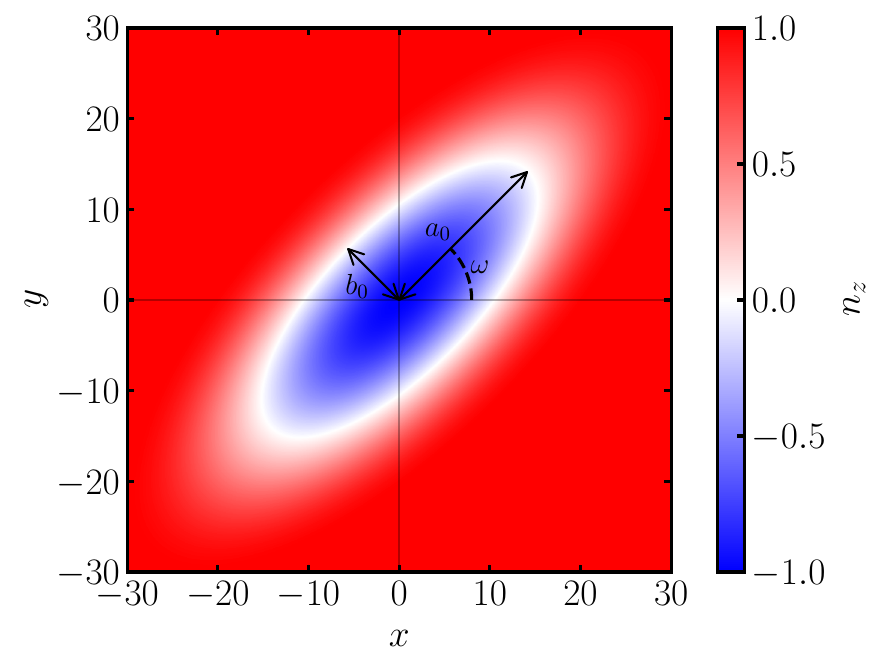}
	\caption{Color map of the $z$-component of Eq.\ \eqref{eq:vectorfield} parametrized via the elliptic triangular approach, highlighting the semi-axes $a_0$, $b_0$, and the rotation angle $\omega$.
		\label{fig:EllipseParameterDiscription}}
\end{figure}

The triangular form of the radial profile reads
\begin{equation}
\begin{split}
\Theta(\rho,\varphi) &= \left(\pi - \xi \right)\mathcal{H}\left(\pi - \xi\right), \quad \text{with~} \xi=\frac{\pi \rho}{2 \rho_0(\varphi)},
\end{split}
\end{equation}
where $\mathcal{H}\left(x\right)$ denotes the Heaviside step function. The $\varphi$-dependence enters through the antiskyrmion radius $\rho_0(\varphi)$, defined as the distance from the skyrmion center to the point where the $z$-component of the magnetization vanishes, i.e., where the magnetization lies in the $x$-$y$-plane. To capture the directional anisotropy of the antiskyrmion, we employ an elliptical ansatz for $\rho_0(\varphi)$, characterized by the semi-axes $a_0$ and $b_0$, and the in-plane rotation angle $\omega$, i.e., 
\begin{equation}
\label{eq:ellipticitiy}
\begin{split}
\rho_0(\varphi) &= \frac{a_0 b_0}{\sqrt{b_0^2 \cos^2(\varphi - \omega) + a_0^2 \sin^2(\varphi - \omega)}}.
\end{split}
\end{equation}
Figure \ref{fig:EllipseParameterDiscription} shows the component $n_z$ of Eq.\ \eqref{eq:vectorfield} using this triangular elliptic parametrization. The plot visualizes the semi-axes $a_0$ and $b_0$, as well as the rotation angle $\omega$ of the elliptic antiskyrmion. 

We point out that in the following derivation, the elliptic triangular approximation enforces the antiskyrmion profile to have a triangular shape at all times, which imposes a fixed relation between the wall width and the radius. It is known that the domain-wall width is determined by the competition between exchange interaction, the DMI and, if present, the anisotropy \cite{Wang,Zhuo}. Furthermore, recent studies have demonstrated that an elliptic deformation of the out-of-plane magnetization component induces a corresponding elliptic deformation of the in-plane component of skyrmions hosted by a curved magnetic film \cite{YershovKravchuk,BichsKravchuk}. 
Numerical simulations, presented in Sec.~\ref{sec:numerics}, indicate indeed weak quantitative effects due to detailed modulations of $\Phi$ in the antiskyrmion. In view of the strongly simplified triangular profile, which e.g.\ ignores the smooth internal structure of the domain wall, incorporating this $\Phi$-modulations would go beyond the pursued overall accuracy of our model, and it would inhibit the closed analytical treatment presented here. Consequently, the framework used in this work should not be interpreted as a quantitatively accurate description of the full internal antiskyrmion structure, but rather as a simplified collective-coordinate approach that captures the qualitative trends of the antiskyrmion's shrinking behavior in a host material with isotropic bulk DMI.

\section{Energy functional of an elliptic triangular antiskyrmion}

Plugging the elliptic triangular ansatz into the energy density in Eq. \eqref{eq:EnergyDensity} yields for the derivatives
\begin{align}
\partial_\rho \Theta&=-\frac{\pi}{2\rho_0} \, , \\
\partial_\varphi \Theta&= \frac{(b_0^2 - a_0^2)\,\pi\,\rho\,\rho_0}{4\,a_0^2 b_0^2}
\sin\!\left(2(\varphi - \omega)\right) \, ,
\end{align}
and the total energy is then given by the integral
\begin{equation}
\begin{split}
E(a_0,e_0,\varphi_0,\omega)=\int_0^{2\pi} d\varphi \int_0^{2\rho_0} d\rho \, \rho\mathcal{E}(a_0,e_0,\varphi_0,\omega,\rho,\varphi).
\end{split}
\end{equation}
While the integration over $\varphi$ cannot be performed analytically, the integral over $\rho$ can be evaluated explicitly and we find 
\newpage
\begin{widetext}
\begin{equation}
\label{eq:totalEnergy}
\begin{split}
E(a_0,b_0,\varphi_0,\omega)=\int_0^{2\pi} d\varphi \, & \left\lbrace J  \left[ 
\frac{1}{4} \left( C + \pi^2 - \operatorname{Ci}\!\left(2\pi\right) + \ln (2\pi) \right) + \frac{ \left( a_0^2 - b_0^2 \right)^2 \pi^2 
\sin^2\!\left( 2\varphi - 2\omega \right)}
{16\left( b_0^2 \cos^2\!\left(\varphi - \omega\right) + a_0^2 \sin^2\!\left(\varphi - \omega\right) \right)^2}
\right] \right. \\
&+ \tilde{D} \frac{ a_0 b_0 \pi \left[ \left(a_0^2 + b_0^2\right)\sin\!\left(2\varphi - \varphi_0\right) 
+ \left(a_0^2 - b_0^2\right)\sin\!\left(\varphi_0 - 2\omega\right) \right]}
{ 2 \left( b_0^2 \cos^2\!\left(\varphi - \omega\right) + a_0^2 \sin^2\!\left(\varphi - \omega\right) \right)^{3/2} } \\
&\left.+ \tilde{B} \frac{ 2 a_0^2 b_0^2 \left( \pi^2 - 4 \right) }
{ \pi^2 \left( b_0^2 \cos^2\!\left(\varphi - \omega\right) + a_0^2 \sin^2\!\left(\varphi - \omega\right) \right) } \right\rbrace .
\end{split}
\end{equation}
\end{widetext}
For $a_0 = b_0$, corresponding to a circular antiskyrmion, the integral can be evaluated, yielding
\begin{equation}
\label{eq:TotalEnergyCircular}
\begin{split}
E(a_0)=&J \left[\frac{\pi}{2}  \left(C + \pi^2 - \operatorname{Ci}(2\pi) + \ln(2\pi)\right)\right] \\
&+ \left(\pi -\frac{4}{\pi} \right) 4 a_0^2 \tilde{B}.
\end{split}
\end{equation}
With the Euler-Mascheroni constant $C$ and the cosine integral $\operatorname{Ci}$ \cite{Gradshteyn}, the expression in the square brackets evaluates to $19.3322$. This energy is similar to the skyrmion case, except that it is independent of the DMI and $\varphi_0$, as known for circular antiskyrmions in a bulk DMI environment \cite{LauDiss}. It is also evident that a circular antiskyrmion has a higher energy than the homogeneous ferromagnet state, since the Zeeman term increases the total energy for finite values of $a_0$.

\begin{figure}[!]
	\centering
	\includegraphics[width=0.47\textwidth]{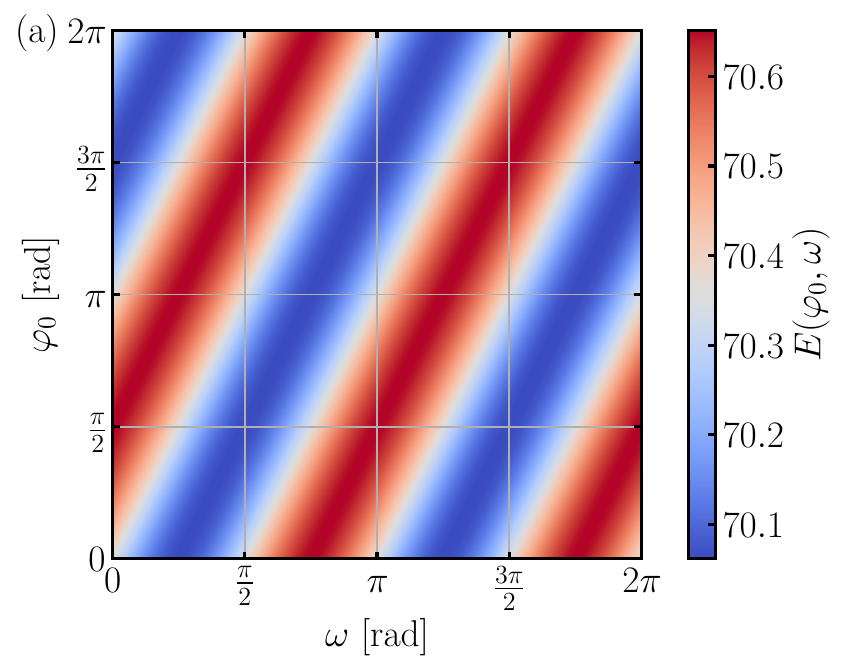}
	\includegraphics[width=0.47\textwidth]{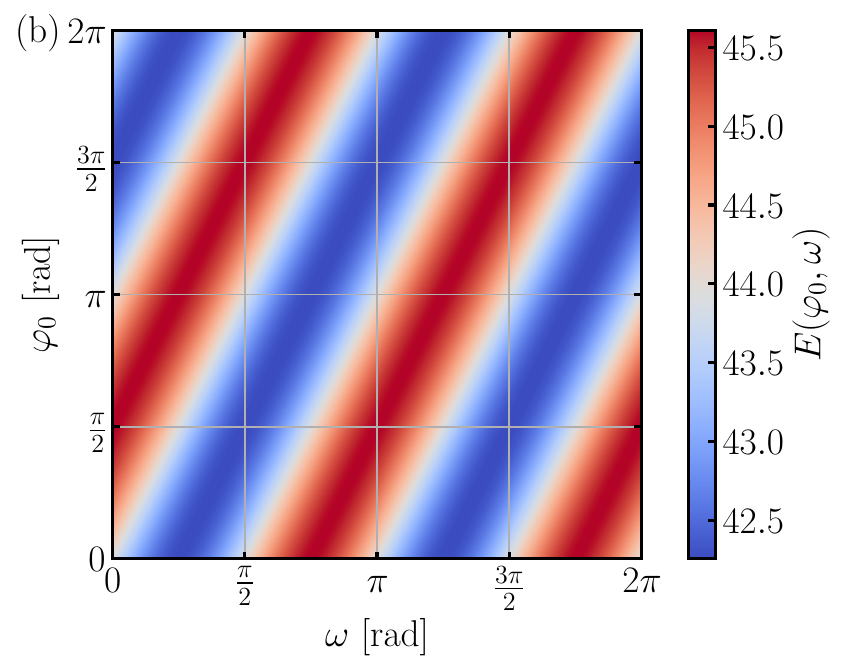}
	\caption{Color maps of the total energy $E(\varphi_0, \omega)$ for elliptic antiskyrmions obtained from numerical integration of Eq.\ \eqref{eq:totalEnergy}. Panel (a) shows the case $\tilde{D}=0.02$, $\tilde{B}=0.02$, $a_0=20$, $b_0=17$, and panel (b) shows $\tilde{D}=0.04$, $\tilde{B}=0.02$, $a_0=15$, $b_0=4$. The color maps illustrate the $\pi$-periodicity of the energy landscape with respect to $\varphi_0$ and $\omega$, with energy minima along paths given by Eq.\ \eqref{eq:minimumEnergyCondition}. Energy maxima occur along lines shifted by $\pi/2$ relative to the minima. While the absolute energies vary with $a_0$, $b_0$, and $\tilde{D}$, the general structure of the color maps remains the same.
		\label{fig:EnergyVarphi0VsOmega}}
\end{figure}

Carrying out the integration in Eq.\ \eqref{eq:totalEnergy} numerically yields color maps of $E(\varphi_0, \omega)$ for different values of $a_0$, $b_0$ and $\tilde{D}$   in Fig. \ref{fig:EnergyVarphi0VsOmega}. While the absolute energy depends on these parameters, the general structure of the color maps stays the same. An energy minimum path goes through $(\varphi_0,\omega)=(\pi/2, \pi/2)$ and $(\varphi_0,\omega)=(3\pi/2, \pi)$, consistent with the $\pi$-symmetry of the ellipse. The energy minima are indicated in blue color and are located at
\begin{equation}
\label{eq:minimumEnergyCondition}
\varphi_0(\omega)=2\omega-(4\ell +1)\frac{\pi}{2}, \quad \ell \in \mathbb{Z} \,  .
\end{equation}
These grooves can be attributed to the DMI term in Eq. \eqref{eq:totalEnergy}, involving an argument
\begin{equation}
\sin(\varphi_0 - 2\omega)=-\sin\left((4\ell +1)\frac{\pi}{2}\right)=-1 \quad \ell \in \mathbb{Z},
\end{equation}
which indicates its significant role in the energy functional.

\begin{figure}[!]
	\centering
	\includegraphics[width=0.47\textwidth]{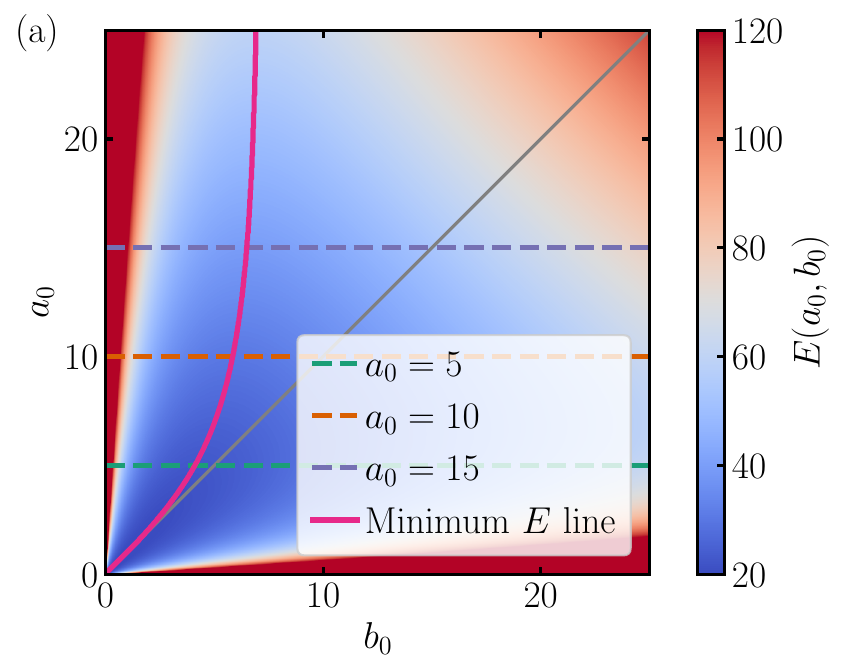}
	\includegraphics[width=0.47\textwidth]{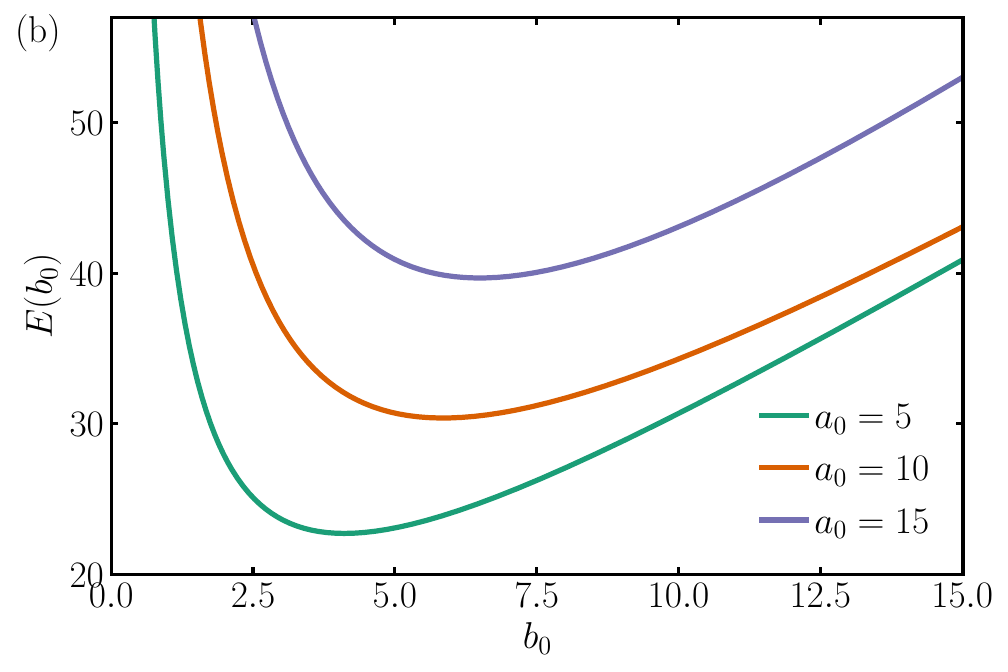}
	\caption{(a) Color map of the total energy $E(a_0, b_0)$ for elliptic antiskyrmions, with cuts indicated by colored dashed lines. (b) Energy along these cuts $E(b_0)$, at fixed values $a_0$ of the major semi-axis. Smaller antiskyrmions generally have lower energy than larger ones. Along cuts for $a_0=$const., the energy exhibits a minimum at $b_0 < a_0$ for a finite $b_0>0$. The magenta line in (a) indicates the minimum energy path on the energy surface, while the gray line corresponds to $a_0 = b_0$. Elliptical antiskyrmions are energetically favorable compared to circular ones. The parameters are $\tilde{D}=0.02$, $\tilde{B}=0.02$, $\varphi_0=0$, and $\omega=\pi/2$.
		\label{fig:Energya0Vse0}}
\end{figure}

Figure \ref{fig:Energya0Vse0} (a) shows a color map of $E(a_0, b_0)$, with slices indicated by colored dashed lines, which are displayed in the $E(b_0)$ plot in (b). We find that smaller antiskyrmions generally have lower energy than bigger ones (though always above the ferromagnetic state). Along slices at fixed values $a_0$ of the semi-axis, there is an energy minimum at $b_0<a_0$ for a finite $b_0>0$. The magenta line in (a) indicates the minimum energy path through the energy surface, while the gray line corresponds to $a_0=b_0$. The minimum energy path (magenta line) can also be mirrored across the gray line. Elliptical antiskyrmions are always energetically favorable compared to circular ones, although at small $a_0$ the behavior deviates from that at large $a_0$, indicating less eccentric elliptic antiskyrmions for smaller values of $a_0$.

\begin{figure}[!]
	\centering
\includegraphics[width=0.47\textwidth]{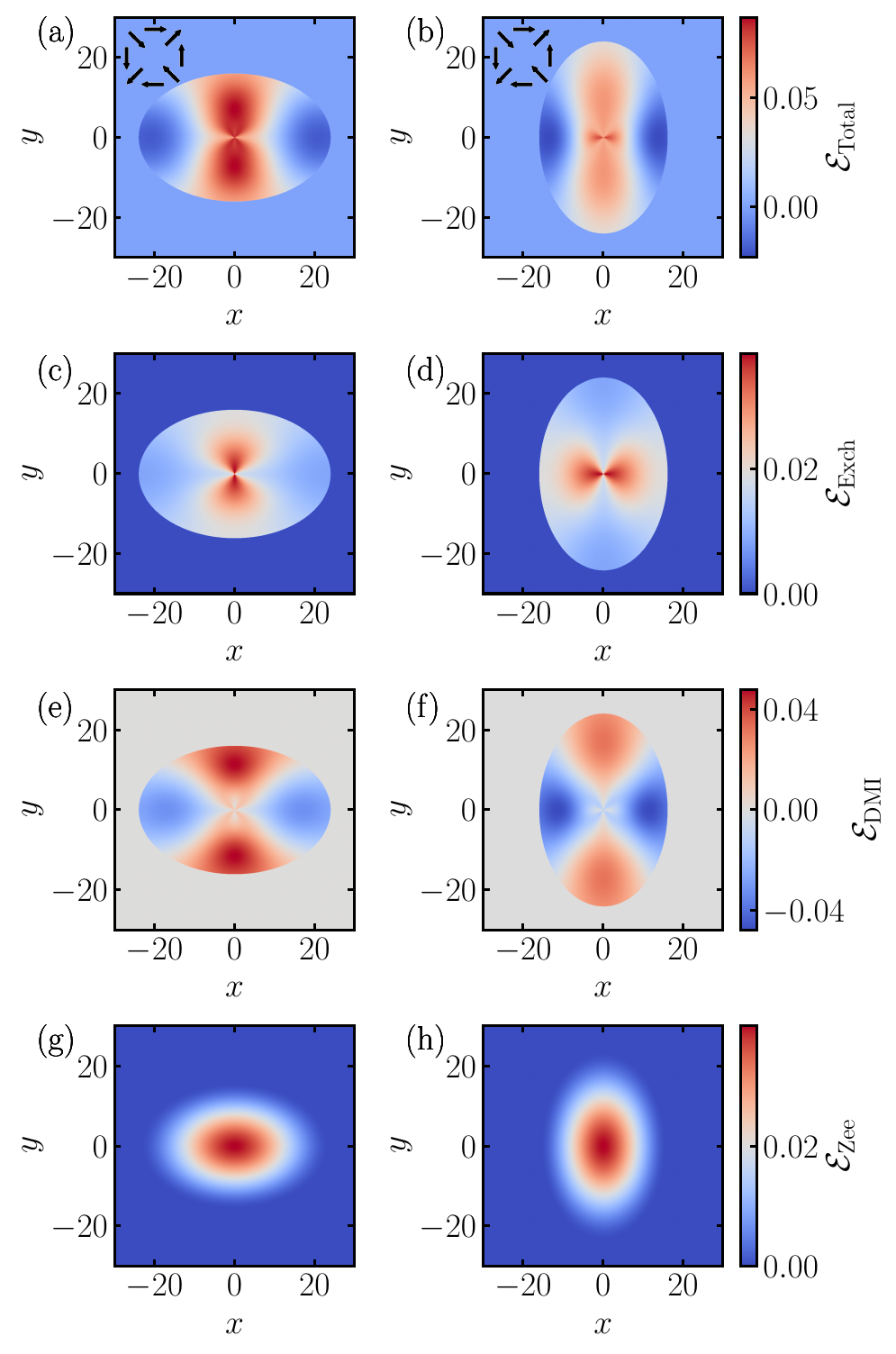}
	\caption{Energy density of an elliptic triangular antiskyrmion for two cases of the in-plane rotation angle, $\omega = 0$ (panels (a), (c), (e), (g)) and $\omega = \pi/2$ (panels (b), (d), (f), (h)). Panels (a) and (b) show the total energy density including all energy contributions, with the vector field configuration indicated by the pictogram ($\varphi_0 = \pi/2$). Panels (c) and (d), (e) and (f), and (g) and (h) show the contributions from the exchange, DMI, and Zeeman terms, respectively. While the exchange and Zeeman contributions are rotationally invariant, the DMI contribution varies with $\omega$, leading to a higher total energy density for $\omega = 0$ compared to $\omega = \pi/2$. Parameters: $\tilde{D}=0.2$, $\tilde{B}=0.02$, $a_0=12$, $b_0=8$, $\varphi_0=\pi/2$.
		\label{fig:EnergydensityComponents}}
\end{figure}

In Fig.\ \ref{fig:EnergydensityComponents}, we show the energy density of the elliptic triangular antiskyrmion for two different case of the in-plane rotation angle, $\omega=0$ and $\omega=\pi/2$. Panels (a) and (b) show the energy density including all terms of the energy contributions, with the pictogram in the upper left corner indicating the vector field configuration given by $\varphi_0=\pi/2$. This illustrates that the energy density is not rotational invariant, as already suggested by Fig. \ref{fig:EnergyVarphi0VsOmega}. Panels (c) and (d) show the contribution from the exchange term, (e) and (f) the DMI term, and (g) and (h) the Zeeman term. It is evident, that the exchange and Zeeman contributions are rotationally invariant, whereas the DMI term is not, leading to a significantly larger contribution to the entire energy density in the $\omega=0$ case as compared to $\omega=\pi/2$. This arises because the favorable Bloch-like winding of the antiskyrmion aligns along the $x$-direction, while the energetically most costly configuration corresponds to the anti-Bloch-like winding along the $y$-direction. The (anti-)Néel-type windings occur in between these orientations and are energetically degenerate, lying intermediate to the two Bloch-type cases. Therefore, the $\omega=\pi/2$ rotation minimizes the area of unfavorable anti-Bloch- and (anti-)Néel-type regions, compared to the $\omega=0$ rotation, as seen in Fig.\  \ref{fig:EnergyVarphi0VsOmega}.

\section{Shrinking dynamics at vanishing DMI}

To describe the shrinking behavior of the antiskyrmion, we assume its center as stationary, not moving, which allows integration over the coordinates $\rho$ and $\varphi$. The time-dependent quantities are the two semi-axes $a_0(t)$ and $b_0(t)$, the helicity $\varphi_0(t)$ and the rotation angle $\omega(t)$. To obtain a set of coupled differential equations, the elliptic triangular ansatz is first inserted into the vector field in Eq.\ \eqref{eq:vectorfield} and subsequently differentiated with respect to time. This yields expressions for the components of $\dot{\bm{n}}$ containing $\dot{a}_0$, $\dot{b}_0$, $\dot{\varphi}_0$, and $\dot{\omega}$. The same ansatz is then inserted into the LLG in Eq.\ \eqref{eq:LLG}, providing another set of equations for the vector field components. By weighting the components with weighting functions and combining them, we obtain isolated equations for the time-dependent parameters. In the absence of DMI, this method yields closed-form analytical expressions, as explicated in Appendix \ref{appendixA}, which we subsequently analyze and solve numerically. 

Here, we rewrite the resulting equations in terms of the dimensionless ratio $\chi=a_0/b_0$ to highlight the structural dependence of the dynamical equations on the semi-axes in the exchange terms. Employing this ratio the equations read
\begin{equation}
\label{eq:dota0chi}
\begin{split}
\dot{a}_0& = \frac{a^2 \alpha}{1+\alpha^2}\left\lbrace -\frac{3}{\pi^2}\tilde{B} a_0 \right. \\
&\left. + \frac{3}{4}\frac{\operatorname{Si}\!\left(2\pi\right)}{8\pi}\frac{J}{a_0}\left[\frac{7}{2}+
\left(1-\frac{8\pi}{\operatorname{Si}\!\left(2\pi\right)}\right)\chi^2-\frac{1}{2}\chi^4\right]\right\rbrace,
\end{split}
\end{equation}
\begin{equation}
\label{eq:dotb0chi}
\begin{split}
\dot{b}_0& = \frac{a^2 \alpha}{1+\alpha^2}\left\lbrace -\frac{3}{\pi^2}\tilde{B} b_0 \right. \\
&\left. + \frac{3}{4}\frac{\operatorname{Si}\!\left(2\pi\right)}{8\pi}\frac{J}{b_0}\left[\frac{7}{2}+
\left(1-\frac{8\pi}{\operatorname{Si}\!\left(2\pi\right)}\right)\frac{1}{\chi^2}-\frac{1}{2\chi^4}\right]\right\rbrace,
\end{split}
\end{equation}
\begin{equation}
\label{eq:dotphi0chi}
\begin{split}
&\dot{\varphi}_0 = \frac{a^2}{1+\alpha^2}\left\lbrace\tilde{B} \vphantom{\frac{3}{2}} \right. \\
&\left. -\frac{\operatorname{Si}\!\left(2\pi\right)}{16\pi}
\frac{\pi^2J}{a_0^2+b_0^2}\left[1-\frac{8\pi}{\operatorname{Si}\!\left(2\pi\right)}+
\frac{3}{2}\left(\chi^2+\frac{1}{\chi^2}\right) \right]\right\rbrace,
\end{split}
\end{equation}
\begin{equation}
\label{eq:dotomega}
\begin{split}
\dot{\omega} =0, \qquad\qquad\qquad\qquad\qquad\qquad\qquad\qquad\qquad\,
\end{split}
\end{equation}
with the sine integral $\operatorname{Si}$ \cite{Gradshteyn}. Note that the pair of Eqs.\ \eqref{eq:dota0chi} and \eqref{eq:dotb0chi} is symmetric under the exchange $a_0\leftrightarrow b_0$ combined with $\chi\leftrightarrow 1/\chi$, as expected.

\subsection{Solution of the coupled differential equations}
\begin{figure}
    \centering
    \includegraphics[width=0.47\textwidth]{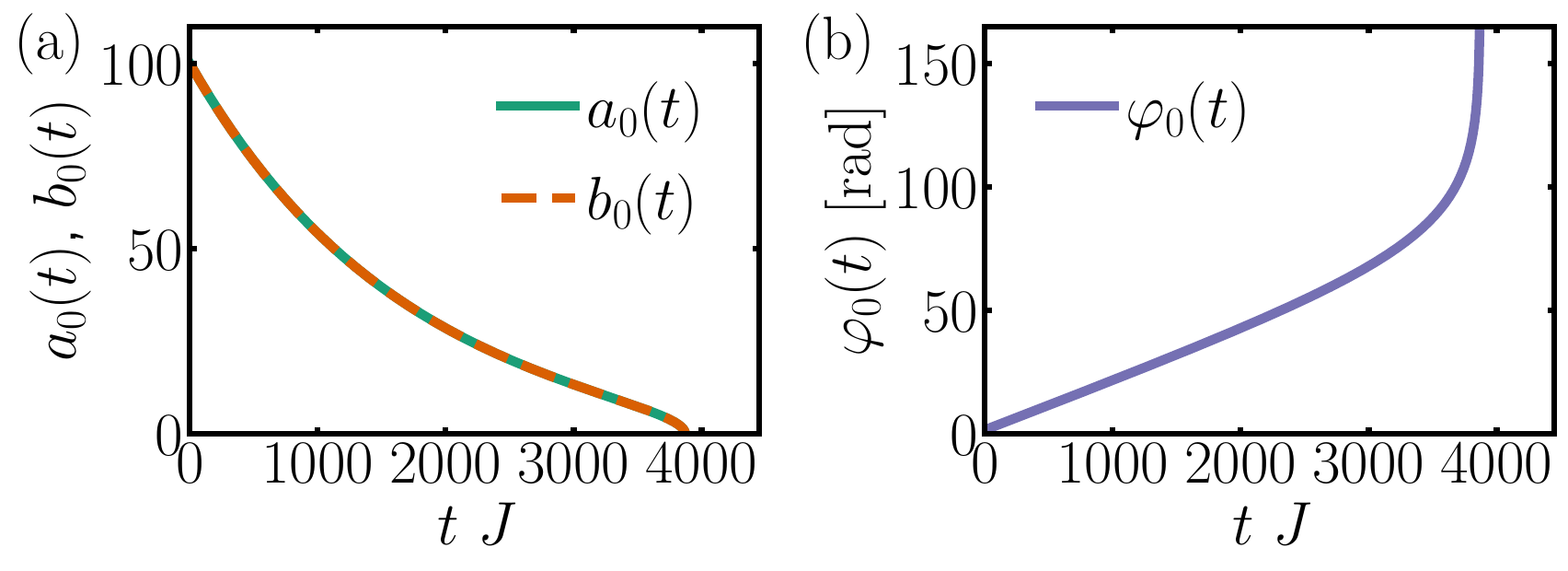}
    \includegraphics[width=0.47\textwidth]{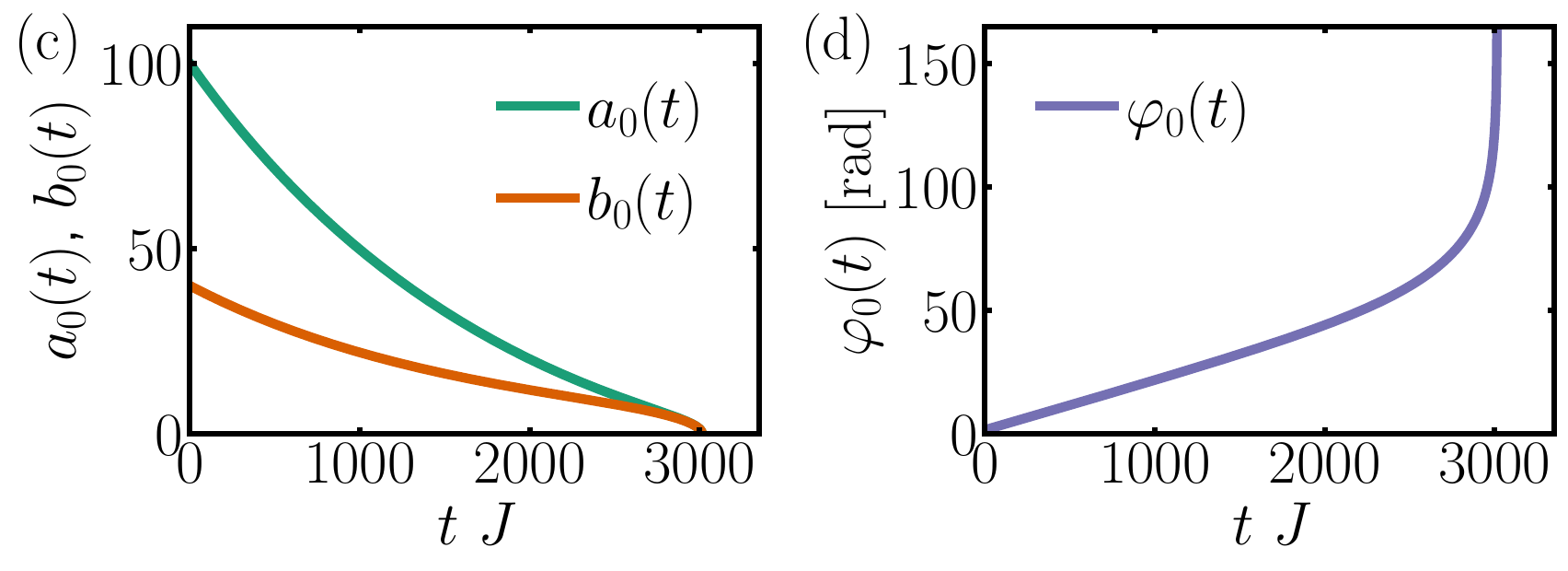}
    \caption{Time evolution of the semi-axes and helicity obtained from the numerical solution of the coupled differential equations following the triangular ansatz.  
Panels (a) and (b) show the circular case with $a_{0,0}=b_{0,0}=100$. Here, the semi-axes dynamical equation reduces to Eq.\ \eqref{eq:dota0circular}, and the helicity follows Eq.\ \eqref{eq:dotphi0circular}.  
Panels (c) and (d) show the elliptical case with $a_{0,0}=100$, $b_{0,0}=40$. Both semi-axes shrink exponentially at large semi-axes with the same rate determined by the Zeeman term. The exchange term dominates at small semi-axes and drives $\chi=a_0/b_0 \to 1$, pushing the system towards the circular limit and the subsequent time evolution follows a square root-like collapse of the semi-axes. The helicity exhibits a linear growth at large semi-axes and a logarithmic divergence as $t\to t_c$ at small semi-axes.  
Parameters: $\tilde{B}=0.02$, $\alpha=0.1$ and $\varphi_{0,0}=\pi/2$.}
    \label{fig:ODESol_D0}
\end{figure}

In the case of a circular antiskyrmion ($a_0=b_0$) the system of coupled differential equations simplifies. The semi-axis dynamics reduce to
\begin{equation}
\label{eq:dota0circular}
 \dot{a}_0 = \frac{a^2 \alpha}{1+\alpha^2} \left[- \frac{3}{\pi^2} \tilde{B}  a_0 - J \frac{1}{a_0} \left( \frac{3}{4} - \frac{3}{8\pi} \operatorname{Si}(2\pi) \right) \right], 
\end{equation}
with $c_1=\tfrac{3}{4}-\tfrac{3}{8\pi}\operatorname{Si}(2\pi)\approx0.58072$. The helicity obeys
\begin{equation}
\label{eq:dotphi0circular}
\begin{split}
\dot{\varphi}_0 =& \frac{a^2}{1 + \alpha^2} \left[ \tilde{B} + \frac{J }{{ a_0^2 }} \left( \frac{ \pi^2}{4}
- \frac{\pi}{8} \operatorname{Si}(2\pi) \right) \right],
\end{split}
\end{equation}
with $c_2=\tfrac{\pi^2}{4}-\tfrac{\pi}{8}\operatorname{Si}(2\pi) \approx1.9105$. Both equations show the same structure found previously for the isotropic skyrmion shrinking case. Figures \ref{fig:ODESol_D0} (a) and (b) displays the time evolution of the equal semi-axes and the helicity, respectively, revealing the same time dependence as known from the solution of the dynamical equations of these structures from Ref.\  \cite{Austrup}.

For an initially elliptical antiskyrmion (\(a_{0,0}>b_{0,0}\)) the Zeeman term dominates at large radii in Eqs.\ \eqref{eq:dota0chi}, \eqref{eq:dotb0chi} and \eqref{eq:dotphi0chi}, and both semi-axes shrink exponentially with the same rate. In that regime
\begin{equation}
a_0(t) \sim a_{0,0}\exp\!\left(-\Gamma t\right), \quad b_0(t) \sim b_{0,0}\exp\!\left(-\Gamma t\right), 
\end{equation}
with the rate \(\Gamma = \dfrac{\alpha a^2}{1+\alpha^2}\dfrac{3}{\pi^2}\tilde B\). In this regime, the helicity increases linearly as
\begin{equation}
\varphi_0(t) \sim \varphi_{0,0} + \frac{a^2}{1+\alpha^2} \tilde{B} t.
\end{equation}
The exchange terms scale inversely with the semi-axes, as $1/a_0$ for the major and $1/b_0$ for the minor axis. Thus, exchange interaction dominates at small semi-axes. In the equation for the helicity, the exchange term scales as $1/(a_0^2+b_0^2)$, making it also the dominant term in the small semi-axes regime.

Figure \ref{fig:ODESol_D0} (c) and (d) display the numerical solution of the coupled nonlinear differential equations for an initially elliptic antiskyrmion. Panel (c) shows the exponential decay of the semi-axes at large radii. As the semi-axes shrink, the exchange term becomes increasingly important. The key to the subsequent dynamics is captured by the evolution of the ratio $\chi$ itself. Computing its time derivative, $\dot{\chi} = (\dot{a}_0 b_0 - a_0 \dot{b}_0)/b_0^2$, by substituting Eqs.\ \eqref{eq:dota0chi} and \eqref{eq:dotb0chi} yields
\begin{equation}
\label{eq:dotchiExch}
\begin{split}
\dot{\chi}=&-\frac{3a^2\alpha}{64\pi\left(1+\alpha^2\right)} \frac{J}{b_0^2} \frac{\left( \chi^2-1 \right)}{\chi^3}\left[\operatorname{Si}\!\left(2\pi\right)\left(1+\chi^4\right) \right. \\
& \left. +2\chi^2\left(8\pi+3\operatorname{Si}\!\left(2\pi\right)\right)\right].
\end{split}
\end{equation}
This result reveals a stationary point at $\chi=1$, where $\dot{\chi}=0$. The stability of this stationary point is confirmed by the derivative evaluated at $\chi=1$ being always negative as
\begin{equation}
\frac{\partial \dot{\chi}}{\partial \chi}\bigg|_{\chi=1}=-\frac{3a^2\alpha\left(2\pi+\operatorname{Si}\!\left(2\pi\right)\right)}{4\pi\left(1+\alpha^2\right)}\frac{J}{b_0^2}<0.
\end{equation}
Therefore, the exchange-driven shrinking at small radii is enhanced for the major axis $a_0$ compared to $b_0$, driving the antiskyrmion towards a circular shape ($\chi \to 1$). Once $a_0=b_0$, the dynamics reduce to those of a circular antiskyrmion, given by Eqs.\ \eqref{eq:dota0circular} and \eqref{eq:dotphi0circular}, leading to a square root-like collapse of the semi-axes
\begin{equation}
\label{eq:squareroot}
a_0(t)=b_0(t) \sim \sqrt{\frac{2 c_1 a^2 \alpha}{1+\alpha^2}J\left(t_c-t\right)},
\end{equation}
with a critical time $t_c$ at which the semi-axes vanish.

Figure \ref{fig:ODESol_D0} (d) shows the corresponding time evolution of the helicity obtained from Eq.\ \eqref{eq:dotphi0chi}. At large radii, $\varphi_0$ increases linearly with time, while at small radii, where $a_0= b_0$, it exhibits a logarithmic divergence, consistent with Eq.\ \eqref{eq:dotphi0circular} and characterized by 
\begin{equation}
\label{eq:logarthmic}
\varphi_0(t) \sim -\frac{c_2}{2c_1} \ln\left(t_c-t\right),
\end{equation}
driving rotation of $\varphi_0$ rapidly.

Thus, for vanishing DMI, the ellipticity is dynamically unstable and any initial ellipticity decreases, such that the system evolves towards the circular shape. Also, in this regime, the dynamics of the semi-axis dynamics is decoupled from that of the helicity. The helicity itself remains dynamically dependent on the semi-axes. Notably, the rotation angle $\omega$ does not enter the dynamics and Eq.\ \eqref{eq:dotomega} suggests that for an initially elliptic antiskyrmion $\omega(t)=\text{const}$.

\vspace{2em}
\section{Shrinking dynamics at finite DMI}

Having established the equations of motion for the exchange and Zeeman terms (i.e., without the DMI), we now turn to the contribution arising from the DMI. The derivation proceeds in complete analogy to the previous section, following the same projection procedure to isolate the individual time-dependent properties. In contrast to the previous case, however, the resulting expressions including the DMI contain square-root factors that prohibit an explicit evaluation of the angular integral over $\varphi$. For this reason, we solve the coupled differential equations of the full system, including exchange, DMI and Zeeman contributions, by direct numerical integration. In a subsequent step, we then apply a Taylor expansion to the DMI term which eventually allows for explicit analytical expressions that reproduce the dynamics with high accuracy. The full DMI expressions as well as the Taylor expansion are shown Appendix \ref{appendixB}.

\subsection{Shrinking of an initially circular antiskyrmion at finite DMI}
\begin{figure*}[!]
    \centering
    \includegraphics[width=\textwidth]{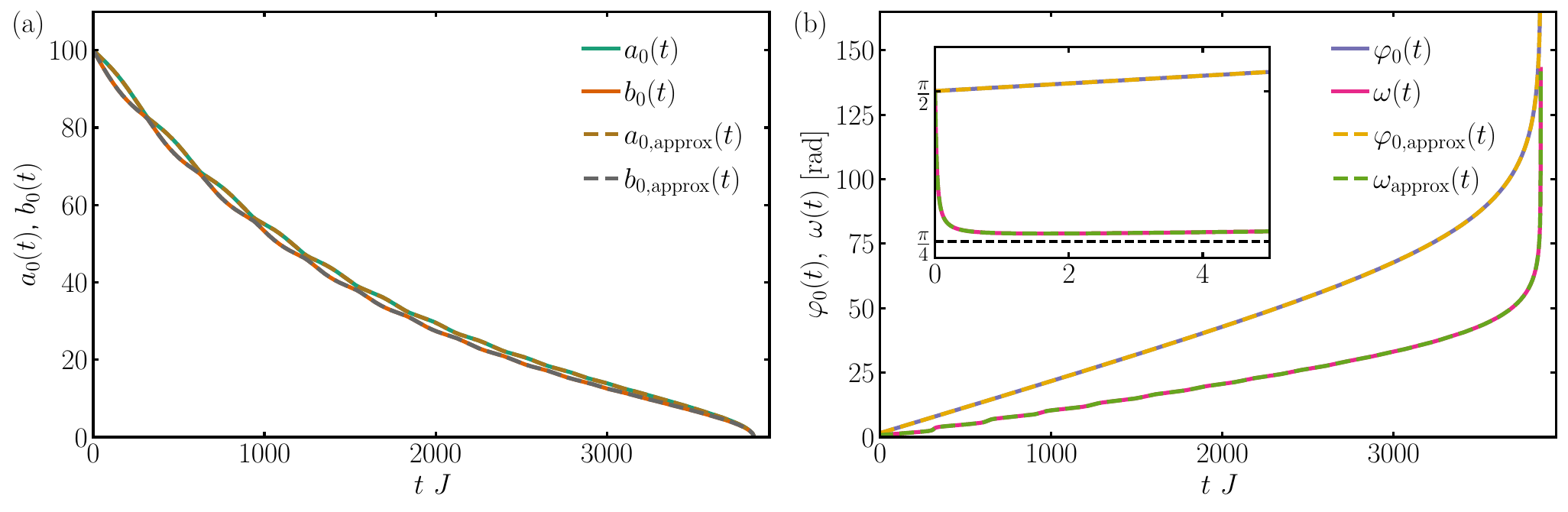}
    \includegraphics[width=\textwidth]{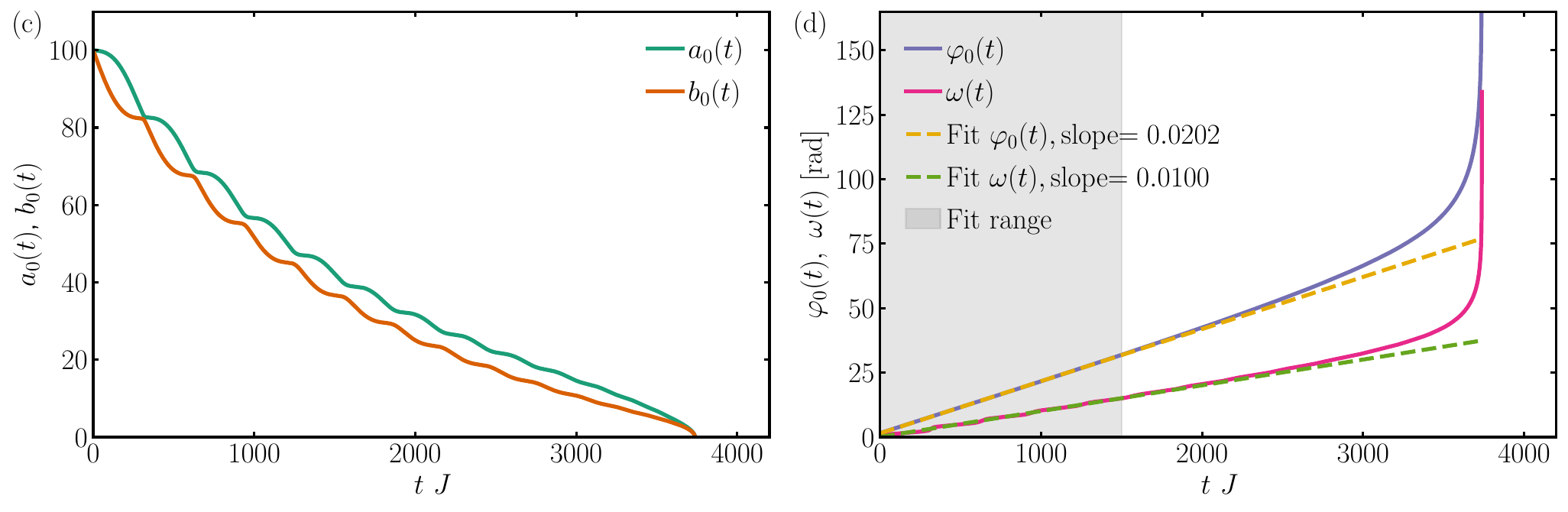}
    \caption{Time evolution of an initially circular antiskyrmion at finite DMI. Panels (a) and (c) show the semi-axes $a_0(t)$ and $b_0(t)$, while panels (b) and (d) display helicity $\varphi_0(t)$ and rotation angle $\omega(t)$ for $\tilde{D}=0.02$ and $\tilde{D}=0.06$, respectively. Dashed lines in (a) and (b) indicate solutions obtained from the approximate DMI contributions (Eqs.\ \eqref{eq:dota0approx}-\eqref{eq:dotomegaapprox}), which agree 
    very well with the full expressions. The inset in (b) highlights the relaxation of $\omega(t)$ from its initial value $\pi/2$ to the equilibrium $\omega_\text{st}$ 
    within very short times, consistent with Eq.\ \eqref{eq:omegaEq}. At large radii the Zeeman term dominates and $a_0(t)$ and $b_0(t)$ decay exponentially, while quadrupole-like oscillations induced by coupling to $\varphi_0$ and $\omega$ due to the DMI become superimposed, with increasing amplitude at larger $\tilde{D}$. Both $\varphi_0(t)$ and $\omega(t)$ evolve linearly in time, with $\omega(t)$ exhibiting step-like jumps synchronized with the quadrupole oscillations. Linear fits in (d) confirm that the slope of $\varphi_0(t)$ scales with the Zeeman field strength $\tilde{B}$ and reveal that $\omega(t)=\varphi_0(t)/2$, reflecting the $2\pi$ rotational symmetry of the helicity versus the $\pi$ symmetry of the ellipse. For small semi-axes, the exchange contribution dominates, driving the system towards circularity and enforcing the square root-like behavior of Eq.\ \eqref{eq:squareroot} for the semi-axes, the logarithmic divergence of Eq.\ \eqref{eq:logarthmic} for $\varphi_0(t)$, and the associated divergence of $\omega(t)$. Parameters: $\tilde{B}=0.02$, $\alpha=0.1$, $a_{0,0}=100$, $b_{0,0}=a_{0,0}-\epsilon$, $\epsilon=10^{-3}$, $\varphi_{0,0}=\pi/2$, $\omega_{0}=\pi/2$.}
    \label{fig:ODESol_circular_finiteD}
\end{figure*}

To investigate whether the system tends towards ellipticity or not, we again introduce the ratio $\chi=a_0/b_0$. Applying this substitution to the approximated DMI contributions in Eqs.\ \eqref{eq:dota0approx} to \eqref{eq:dotomegaapprox} yields
\begin{equation}
\label{eq:dota0DMIchi}
\begin{split}
\dot{a}_{0,\text{approx}}^{\text{DMI}} =& \frac{3 \tilde{D} a^2}{32 \sqrt{2} \pi \sqrt{1+\chi^2}  (1+\alpha^2)} \times \\
& \left[ 4 (\chi^4 + 4\chi^2 + 11) \cos(\varphi_0 - 2\omega) \right. \\
& \left. - (5\chi^4 + 10\chi^2 + 17) \alpha \sin(\varphi_0 - 2\omega) \right],
\end{split}
\end{equation}
\begin{equation}
\label{eq:dotb0DMIchi}
\begin{split}
\dot{b}_{0,\text{approx}}^{\text{DMI}} =& \frac{3 \tilde{D} a^2}{32 \sqrt{2} \pi \chi^3 \sqrt{1+\chi^2} (1+\alpha^2)} \times \\
& \left[ -4 (11\chi^4 + 4\chi^2 + 1) \cos(\varphi_0 - 2\omega) \right. \\
& \left. +  (17\chi^4 + 10\chi^2 + 5) \alpha \sin(\varphi_0 - 2\omega) \right],
\end{split}
\end{equation}
\begin{equation}
\label{eq:dotphi0DMIchi}
\begin{split}
\dot{\varphi}_{0,\text{approx}}^{\text{DMI}} =& -\frac{\tilde{D} \pi a^2(\chi^2 - 1)}{8 \sqrt{2} b_0 \chi \sqrt{1+\chi^2} (1+\alpha^2)} \times \\
&\left[ 10\alpha \cos(\varphi_0 - 2\omega) + 3 \sin(\varphi_0 - 2\omega) \right],
\end{split}
\end{equation}
\begin{equation}
\label{eq:dotomegaDMIchi}
\begin{split}
\dot{\omega}_\text{approx}^{\text{DMI}} =& \frac{3 \tilde{D}a^2}{16 \sqrt{2} \pi b_0 \chi (\chi^2 - 1) \sqrt{1+\chi^2} (1+\alpha^2)} \times \\
& \left[ \alpha (9\chi^4 + 14\chi^2 + 9) \cos(\varphi_0 - 2\omega) \right. \\
& \left. + 16 (1+\chi^2)^2 \sin(\varphi_0 - 2\omega) \right].
\end{split}
\end{equation}
The equation for $\dot{\chi}$ in the presence of DMI acquires now the additional contribution
\begin{equation}
\label{eq:dotchiDMI}
\begin{split}
\dot{\chi}^\text{DMI} =& \frac{3 a^2}{32 \sqrt{2} \pi (1+\alpha^2)} \frac{\tilde{D}}{b_0} \frac{\sqrt{1 + \chi^2}}{\chi^2} \times \\
&\left[ 4\left(1 + 14\chi^2 + \chi^4\right) \cos\!\left(\varphi_0 - 2\omega\right) \right. \\
&\left. - \left(5 + 22\chi^2 + 5\chi^4 \right) \alpha \sin\!\left(\varphi_0 - 2\omega\right) \right].
\end{split}
\end{equation}
In contrast to the zero DMI case $\dot{\chi}$ evaluated at $\chi=1$ reveals that there is no stationary point at $a_0=b_0$
\begin{equation}
\begin{split}
\dot{\chi}|_{\chi=1}=&-\frac{3a^2}{\pi\left(1+\alpha^2\right)} \frac{\tilde{D}}{b_0} \left[ 2\cos\!\left(\varphi_0 - 2\omega\right) \right. \\
& \left. - \alpha \sin\!\left(\varphi_0 - 2\omega\right) \right] \neq 0.
\end{split}
\end{equation}
Therefore, the DMI term provides a finite driving force for $\chi=1$ away from the circular antiskyrmion configuration. Sign and magnitude of this force depend on the DMI strength, the helicity, the rotation angle and the Gilbert damping. This raises the question of how an initially circular antiskyrmion, prepared with a given helicity, tends to align its major axis when it becomes elliptic. The equation for $\dot{\omega}_\text{approx}^\text{DMI}$ in Eq.\ \eqref{eq:dotomegaDMIchi} reaches stationarity $\omega_{\text{st}}$, when the bracket vanishes, leading to the condition
\begin{equation}
\label{eq:omegaEq}
\begin{split}
\omega_{\text{st}}=&\frac{1}{2}\left[\varphi_0
+ \arctan\!\left( \frac{\alpha \left(9+14\chi^2+9\chi^4 \right)}{16\left(1+\chi^2\right)^2} \right) \right. \\
&\left. \vphantom{ \frac{\left(\chi^4\right)}{\left(\chi^2\right)^2}} - \lambda\pi\right],
\quad \lambda \in \mathbb{Z} .
\end{split}
\end{equation}
Figure \ref{fig:ODESol_circular_finiteD} shows the time evolution of the coupled equations of motion obtained from numerical integration. To this end, Eqs.\ \eqref{eq:dota0chi}--\eqref{eq:dotomega} are solved including the DMI contributions in Eqs.\ \eqref{eq:dota0DMI}--\eqref{eq:dotomegaDMI}. The latter are integrated numerically over $\varphi$ at each time step. To avoid the divergence in Eq.\ \eqref{eq:dotomegaDMI} at $\chi=1$, a small asymmetry must be introduced for the circular case, i.e., $b_0=a_0-\epsilon$ with $\epsilon$ being a small parameter.

The time evolution of the semi-axes of an initially circularly prepared antiskyrmion is displayed in (a) and (c) and the time evolution of its helicity and rotation angle in (b) and (d), for $\tilde{D}=0.02$ and $\tilde{D}=0.06$, respectively. For all solutions the initial values are $\varphi_0=\pi/2$ and $\omega=\pi/2$.

Additionally, the solution of the full system is shown by using the approximate DMI contributions of Eqs. \eqref{eq:dota0approx}--\eqref{eq:dotomegaapprox} in panels (a) and (b) as dashed lines. These match the solution of the full expressions with high accuracy. The inset in (b) shows, that the antiskyrmion rotates within very short times away from its initially prepared rotation angle $\omega=\pi/2$ to a value slightly above $\pi/4$. Evaluating Eq.\ \eqref{eq:omegaEq} for initial conditions $\varphi_0=\pi/2$, $\chi \approx 1$, $\alpha=0.1$ and $\lambda=0$ yields $\omega_\text{st}\approx\pi/4+0.05$, in agreement with the observed behavior in (b).

The general time evolution of the semi-axes in panels (a) and (c) follows the exponential decay at large radii. This shrinking is superimposed by quadrupole-like oscillations induced by the coupling between helicity and rotation angle. The amplitude of these oscillations increases with increasing DMI strength. The oscillatory behavior of the semi-axes is directly linked to the rotation of $\varphi_0$ and $\omega$ displayed in (b) and (d). At large semi-axes, both $\varphi_0(t)$ and $\omega(t)$ evolve linearly in time, with $\omega(t)$ showing step-like jumps in phase with the quadrupole oscillations. Linear fits (dashed lines in panel (d)) confirm that the slope of $\varphi_0(t)$ corresponds to the Zeeman field strength $\tilde{B}$, consistent with Eq.\ \eqref{eq:dotphi0chi}, while $\omega(t)=\varphi_0(t)/2$. This reflects the $\pi$ rotational symmetry of the ellipse, as already evident in the energy considerations in Fig.\ \ref{fig:EnergyVarphi0VsOmega}.

The origin of the quadrupole oscillations can be understood by considering the interplay between the elliptic deformation of the antiskyrmion and the coupled dynamics of $\varphi_0(t)$ and $\omega(t)$. Notably, the precession of $\varphi_0(t)$, driven by the external magnetic field, is always present. 

\begin{figure*}[!]
    \centering
    \includegraphics[width=\textwidth]{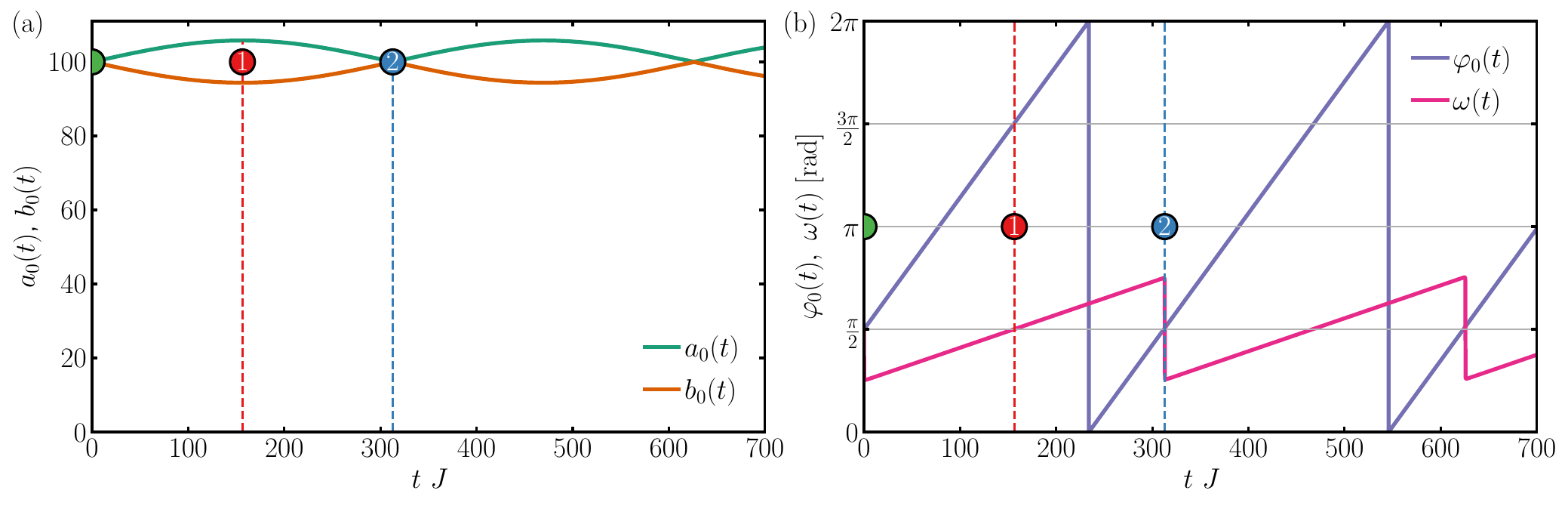}
    \caption{Time evolution of the coupled ODEs in the undamped limit ($\alpha = 0$). Panel (a) shows quadrupole oscillations of the semi-axes $a_0(t)$ and $b_0(t)$ conserving the averaged radius. Panel (b) displays the linear precession of $\varphi_0(t)$, with $\omega(t)$ following the helicity at one quarter of its slope. The green dot marks the start of the evolution, while red and blue dots indicate the time points of most elliptic and circular configurations, respectively. Parameters: $\tilde{D}=0.06$, $\tilde{B}=0.02$, $\alpha=0$, $a_{0,0}=100$, $b_{0,0}=a_{0,0}-\epsilon$, $\epsilon=10^{-3}$, $\varphi_{0,0}=\pi/2$, $\omega_{0}=\pi/2$.}
    \label{fig:ODEQuadrupole_alpha0}
\end{figure*}

In the undamped limit ($\alpha = 0$), the quadrupole oscillations can be directly attributed to energy conservation. Figure  \ref{fig:ODEQuadrupole_alpha0} shows the undamped time evolution of the system obtained from the coupled ODEs. In panel (a), quadrupole oscillations of the semi-axes are observed, conserving the value of the averaged radius. Panel (b) shows the expected linear time evolution of $\varphi_0(t)$, where the data are wrapped within the interval $\left[0,2\pi\right]$. Here, it is revealed that $\omega(t)$ follows the helicity in the undamped regime with one quarter of its slope. The green dot marks the beginning of the time evolution and the red and blue dots mark time points of most elliptic and circular configurations, respectively. As soon as the antiskyrmion becomes circular at (2), $\omega(t)$ exhibits a jump, realigning its rotation angle to a value consistent with Eq.\ \eqref{eq:omegaEq}.

To analyze this behavior further, the corresponding trajectory of $\varphi_0(t)$ and $\omega(t)$ is drawn into the $E(\varphi_0,\omega)$ landscape in Fig.\ \ref{fig:E(phi0,omega)_alpha0} with the same time point indicators as in Fig.\ \ref{fig:ODEQuadrupole_alpha0}. The magnitudes of $E(\varphi_0,\omega)$ are given in arbitrary units, since the absolute energy value depends on the instantaneous magnitudes of the semi-axes and thus varies with time. In general, the $E(\varphi_0,\omega)$ landscape becomes increasingly flat as the ellipticity decreases. However, the qualitative shape of the energy landscape remains unchanged as long as the antiskyrmion is not perfectly circular. In the latter limit, ($a_0 = b_0$), Eq.\ \eqref{eq:TotalEnergyCircular} holds and implies $E(\varphi_0,\omega) = \text{const}$. The plot reveals that the system evolves along a closed path of constant energy. Initially (green dot), the antiskyrmion is circular and becomes elliptic, with its major axis aligning approximately along $\omega \approx \pi/4$ as predicted by Eq.\ \eqref{eq:omegaEq}. The elliptic deformation decreases the system's energy as compared to the circular state. Since energy cannot dissipate when $\alpha=0$, this energy is transferred to rotation in the $E(\varphi_0,\omega)$ landscape towards regions of higher energy. As $\varphi_0$ continues to precess under the external field, $\omega$ and the semi-axes must adapt to maintain constant energy. The antiskyrmion reaches maximal ellipticity when $\omega$ rotates into a maximum in $E(\varphi_0,\omega)$ (1). As $\omega$ continues to drive the rotation, the antiskyrmion evolves back towards the circular state (2). Here, the energy landscape flattens, allowing $\omega$ to realign freely. Consequently, the process repeats periodically. 

\begin{figure}[!]
    \centering
    \includegraphics[width=0.47\textwidth]{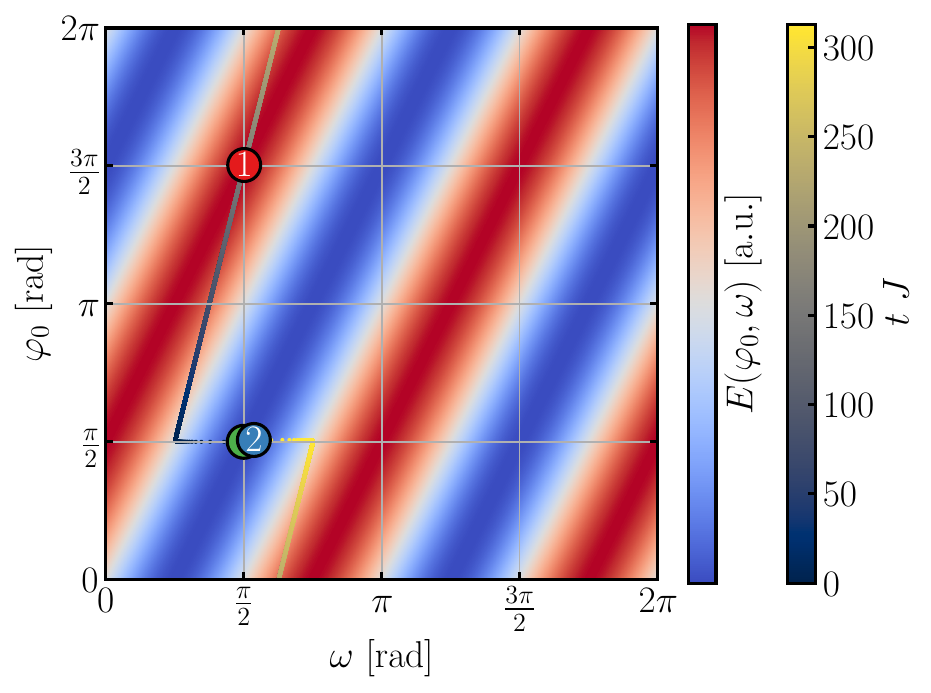}
    \caption{Energy landscape $E(\varphi_0,\omega)$ in the undamped limit ($\alpha = 0$) with the trajectory of $\varphi_0(t)$ and $\omega(t)$ and the time-point indicators from Fig. \ref{fig:ODEQuadrupole_alpha0} (b). The system evolves along a closed, constant-energy path connecting elliptic (1) and circular (2) configurations, starting from the initial state (green dot). In the circular limit, the energy landscape flattens, allowing $\omega$ to realign freely. Parameters: $\tilde{D}=0.06$, $\tilde{B}=0.02$, $\alpha=0$, $a_{0,0}=100$, $b_{0,0}=a_{0,0}-\epsilon$, $\epsilon=10^{-3}$, $\varphi_{0,0}=\pi/2$, $\omega_{0}=\pi/2$.}
    \label{fig:E(phi0,omega)_alpha0}
\end{figure}

\begin{figure*}[!]
    \centering
    \includegraphics[width=\textwidth]{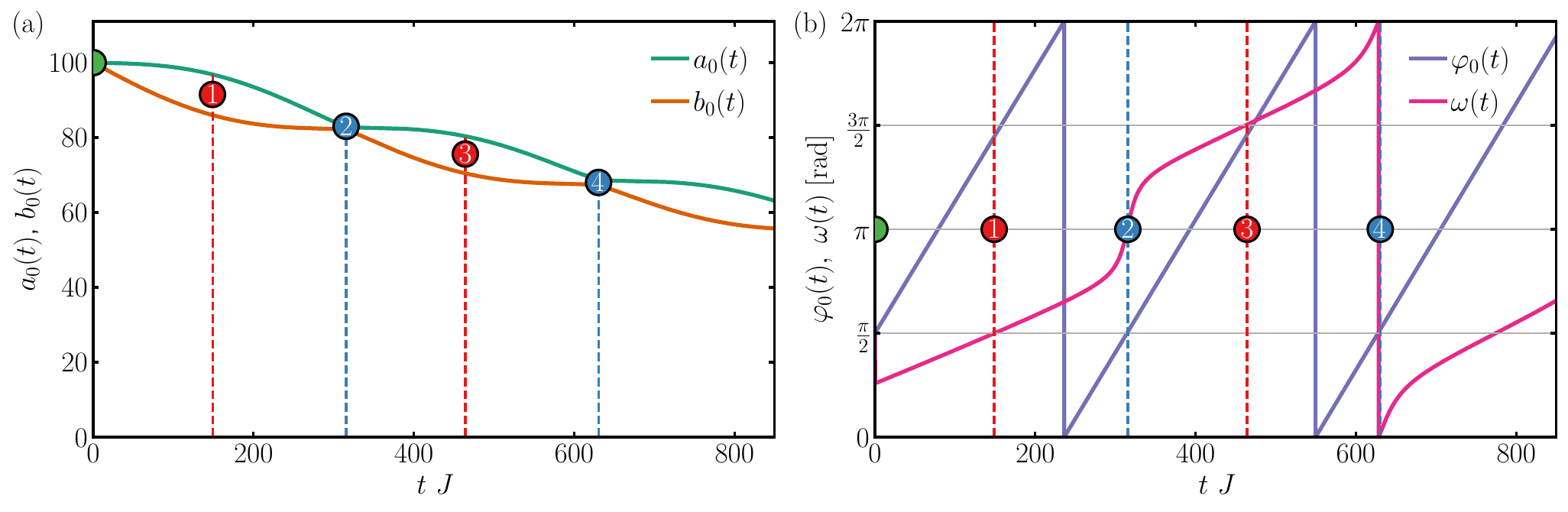}
    \caption{Time evolution of the coupled ODEs in the damped regime ($\alpha=0.1$). Panel (a) displays quadrupole oscillations of the semi-axes accompanied by exponential decay of the averaged radius. Panel (b) shows the corresponding evolution of $\varphi_0(t)$ and $\omega(t)$, with green, red, and blue dots indicating initial, maximal elliptic, and minimal elliptic time points, respectively. Parameters: $\tilde{D}=0.06$, $\tilde{B}=0.02$, $\alpha=0.1$, $a_{0,0}=100$, $b_{0,0}=a_{0,0}-\epsilon$, $\epsilon=10^{-3}$, $\varphi_{0,0}=\pi/2$, $\omega_{0}=\pi/2$.}
    \label{fig:ODEQuadrupole_alpha0.1}
\end{figure*}

For finite damping ($\alpha = 0.1$), the overall mechanism remains similar but energy dissipation alters the behavior. Dissipation is dominated by the Zeeman term at large semi-axes in Eqs.\ \eqref{eq:dota0chi}--\eqref{eq:dotphi0chi}. Thus, energy dissipates primarily through the exponential decay of the averaged radius. Figure \ref{fig:ODEQuadrupole_alpha0.1} focuses on the first oscillations of the antiskyrmion shrinking that is displayed in Fig.\ \ref{fig:ODESol_circular_finiteD} (c) and (d). The green dot marks the beginning of the time evolution and red and blue dots mark time points of maximal and minimal ellipticity, respectively. In contrast to the undamped case, the antiskyrmion does not become perfectly circular during the quadrupole oscillations. Also, $\omega(t)$ does not jump back to the value close to $\pi/4$. 

This can be understood by means of Fig.\ \ref{fig:E(phi0,omega)_alpha0.1} (a). Here, the trajectory of $\varphi_0(t)$ and $\omega(t)$ in the energy landscape $E(\varphi_0,\omega)$ is drawn, with the time point markers corresponding to those of Fig.\ \ref{fig:ODEQuadrupole_alpha0.1}. The energy gain associated with the elliptic deformation is still compensated by a rotation of $\omega$, driven by the precession of $\varphi_0$. However, after maximal ellipticity is reached, the antiskyrmion no longer relaxes back to a perfectly circular state. As a result, $E(\varphi_0,\omega)$ does not flatten, and the system cannot reorient towards $\omega \approx \pi/4$. Returning to this configuration would require passing through an energy minimum, that the antiskyrmions semi-axes would need to compensate for dynamically by increasing the averaged radius. Instead, the antiskyrmion rotates towards $\omega \approx 5\pi/4$, consistent with Eq.\ \eqref{eq:omegaEq} for $\lambda = -2$. This explains the step-like increases observed in $\omega(t)$, where the antiskyrmion briefly accelerates rotation while traversing the energy barrier, whereas the mean radius continues to decay exponentially. For long times, the oscillations of the semi-axes decrease in amplitude and they keep their ellipticity (see Fig.\ \ref{fig:ODESol_circular_finiteD} (c)), and both $\omega(t)$ and $\varphi_0(t)$ evolve near the energy maximum, as shown in Fig.\ \ref{fig:E(phi0,omega)_alpha0.1} (b), where the whole trajectory of $\varphi_0(t)$ and $\omega(t)$ of Fig.\ \ref{fig:ODESol_circular_finiteD} (d) is displayed.

\begin{figure}[!]
    \centering
    \includegraphics[width=0.47\textwidth]{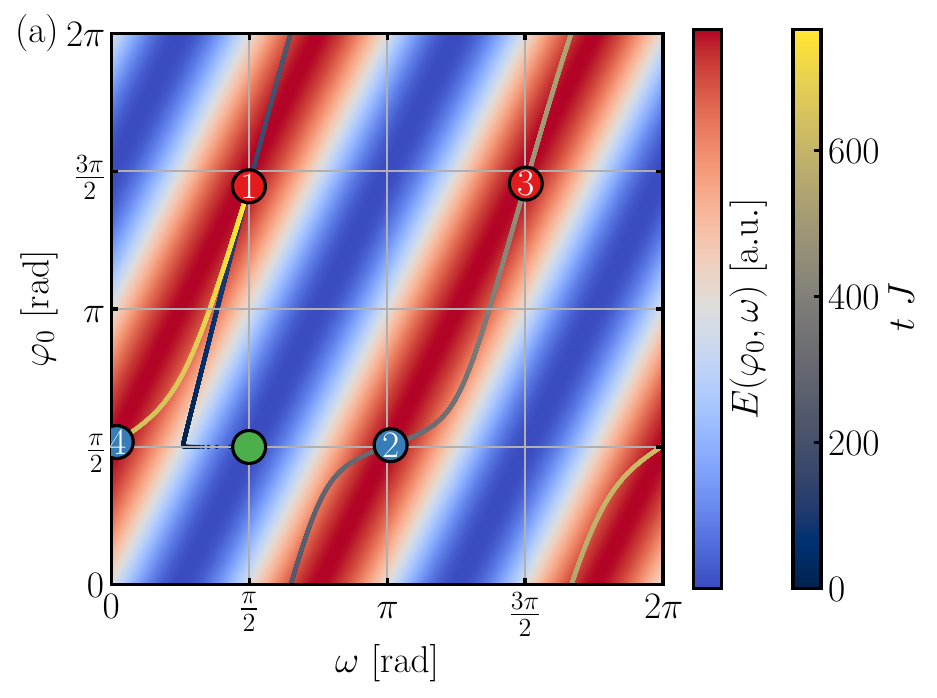}
    \includegraphics[width=0.47\textwidth]{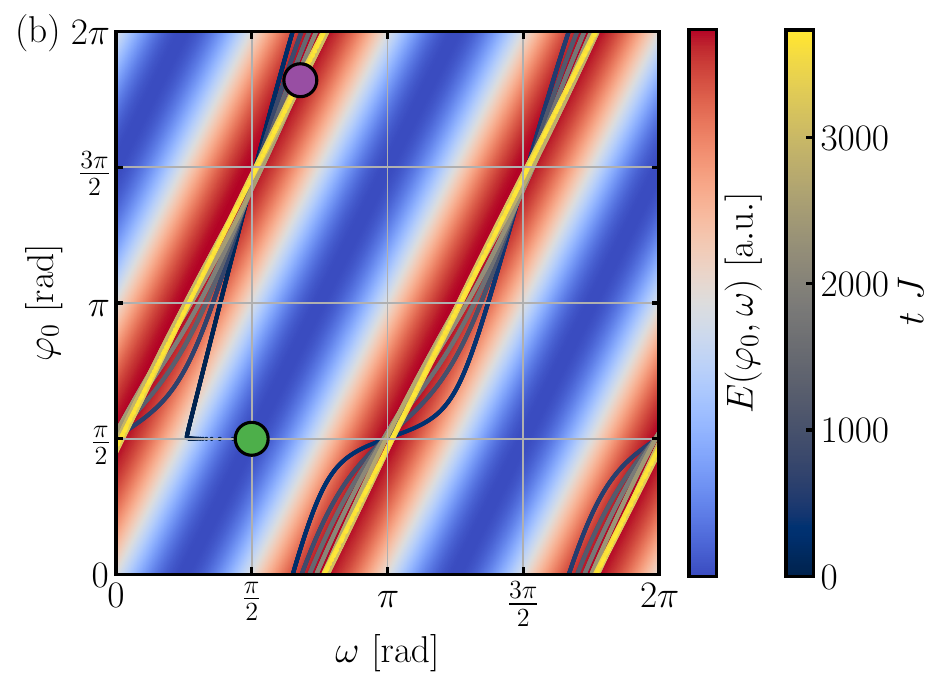}
    \caption{Energy landscape $E(\varphi_0,\omega)$ for finite damping ($\alpha=0.1$). Panel (a) shows the trajectory of $\varphi_0(t)$ and $\omega(t)$ during the first oscillation periods, showing that the energy gain from elliptic deformation is compensated by a rotation of $\omega$. In contrast to the undamped case, the system no longer relaxes to a circular state after maximal ellipticity, instead the trajectory shifts towards $\omega \approx 5\pi/4$, consistent with Eq.\ \eqref{eq:omegaEq} for $\lambda=-2$. Panel (b) displays the full trajectory corresponding to Fig. \ref{fig:ODESol_circular_finiteD}(d), illustrating how it remains near the energy maximum while the oscillation amplitude of the semi-axes decays over time and the antiskyrmion keeps ellipticity. Parameters: $\tilde{D}=0.06$, $\tilde{B}=0.02$, $\alpha=0.1$, $a_{0,0}=100$, $b_{0,0}=a_{0,0}-\epsilon$, $\epsilon=10^{-3}$, $\varphi_{0,0}=\pi/2$, $\omega_{0}=\pi/2$.}
    \label{fig:E(phi0,omega)_alpha0.1}
\end{figure}

In summary, the system primarily dissipates energy through the shrinking of the averaged radius, while the quadrupole oscillations and the rotation of $\omega$ act to compensate the continuous $\varphi_0$ precession imposed by the external field. This compensation arises because, due to the twofold rotational symmetry of the elliptic antiskyrmion, its energy explicitly depends on $\varphi_0$ for fixed $\omega$.

The differential equation for the ratio of the semi-axes, $\dot{\chi}$, scales as $1/b_0^2$ in the exchange term in Eq.\ \eqref{eq:dotchiExch} and as $1/b_0$ in the DMI term in Eq.\ \eqref{eq:dotchiDMI}. Consequently, at small semi-axes the exchange contribution dominates the dynamics of $\chi$ and also the dynamics of the overall system. In this regime, the antiskyrmion evolves towards a circular configuration and $a_0(t)$ and $b_0(t)$ cross over from exponential to square root decay according to Eq.\ \eqref{eq:squareroot}, while $\varphi_0(t)$ diverges logarithmically as described by Eq.\ \eqref{eq:logarthmic}. The rotation angle $\omega(t)$ follows the logarithmic divergence of the helicity, such that both $\varphi_0$ and $\omega$ accelerate rapidly as the semi-axes vanish.

\subsection{Shrinking of an initially elliptic antiskyrmion at finite DMI}

\begin{figure}[!]
    \centering
    \includegraphics[width=0.47\textwidth]{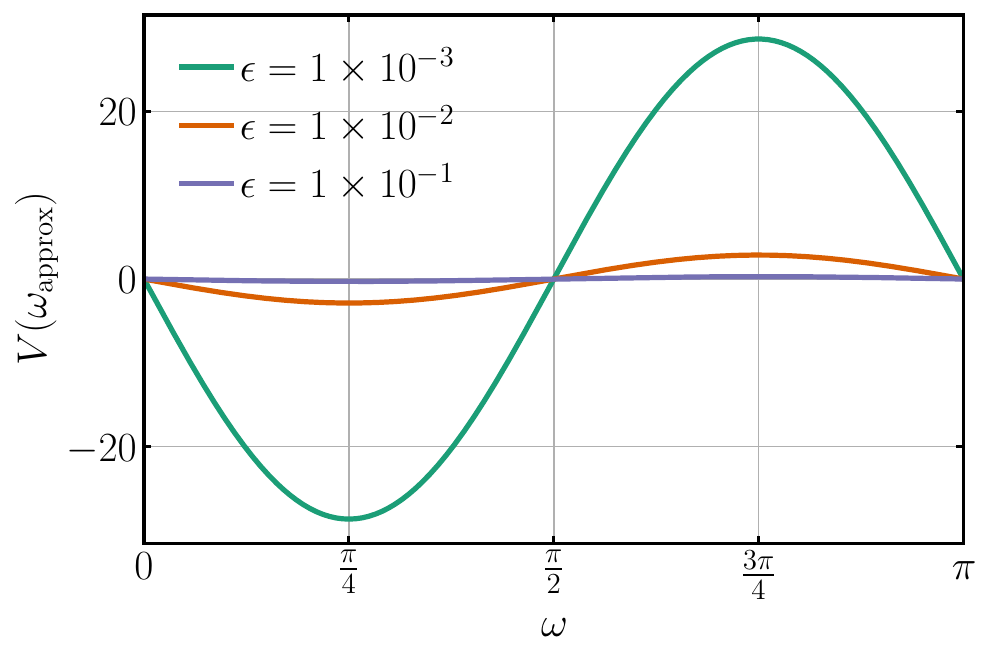}
    \caption{Potential $V(\omega)$ for increasing values of $\epsilon$. The overall amplitude is strongly suppressed with increasing $\epsilon$. Consequently, the torque aligning $\omega$ vanishes rapidly once the initial shape is elliptic, effectively pinning $\omega$ in the early stages of the dynamics. Parameters: $\tilde{D}=0.06$, $\tilde{B}=0.02$, $a_0=100$, $b_0=a_0-\epsilon$, and $\varphi_0=\pi/2$.}
    \label{fig:V(omega)}
\end{figure}

The orientational dynamics of the antiskyrmion is determined by an effective potential $V(\omega)$, defined via $\dot{\omega}=-\partial V(\omega)/\partial \omega$ and obtained from  Eq.\ \eqref{eq:dotomegaDMIchi}. This potential reveals how the torque acting on the rotation angle evolves as the antiskyrmion becomes increasingly elliptical, i.e., with increasing $\chi$. Its explicit form is
\begin{equation}
\begin{split}
\label{eq:V(omega)}
V(\omega)=&\frac{3 a^2 }
{32 \sqrt{2} \pi (1+\alpha^2)} \frac{\tilde{D}}{b_0} \frac{1}{\chi \left( \chi^2-1 \right) \sqrt{1+\chi^2}} \times \\ 
&\left[\left(9 + 14 \chi^2 + 9 \chi^4 \right) \alpha \sin(\varphi_0-2\omega) \vphantom{\left(1+\chi^2\right)^2} \right. \\
& \left. -16\left(1+\chi^2\right)^2\cos(\varphi_0-2\omega)
\right].
\end{split}
\end{equation}
Figure \ref{fig:V(omega)} illustrates $V(\omega)$ for increasing values of $\epsilon$. The overall amplitude of the potential decreases notably for small increases of $\epsilon$, i.e., for increasing $\chi$. Consequently, the torque aligning the rotation angle vanishes rapidly once the initial shape is elliptic, where the small amplitude of the potential $V(\omega)$  effectively pins $\omega$ in the early stages of the dynamics.

\begin{figure*}[!]
    \centering
    \includegraphics[width=\textwidth]{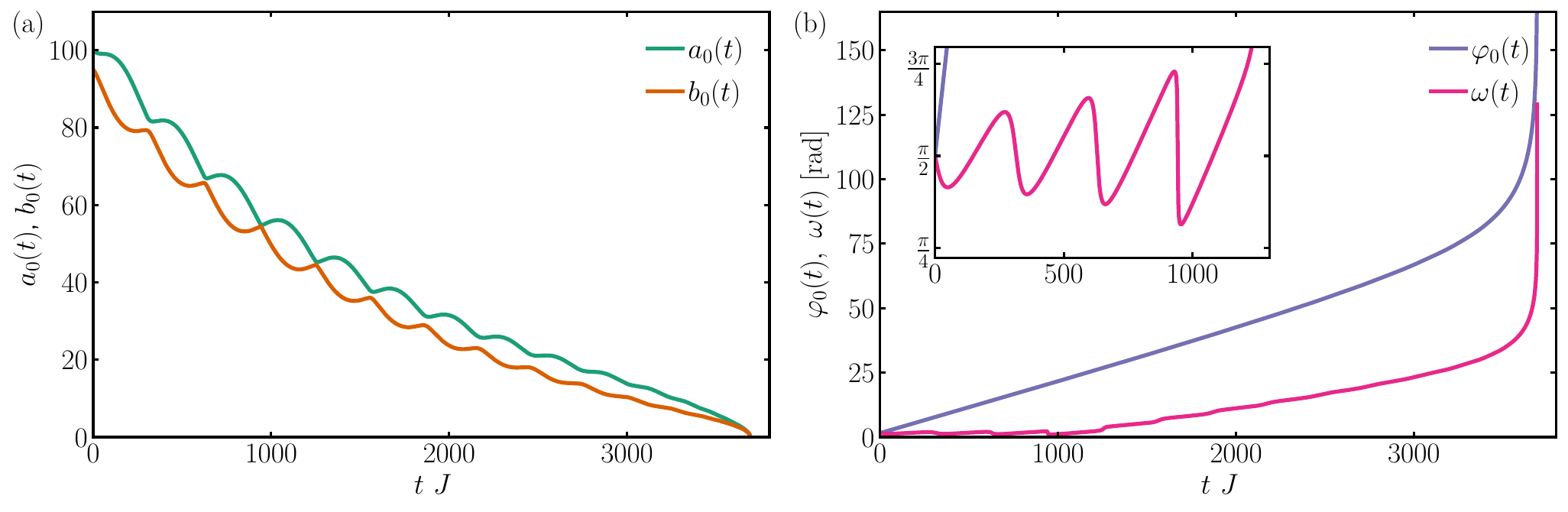}
    \includegraphics[width=\textwidth]{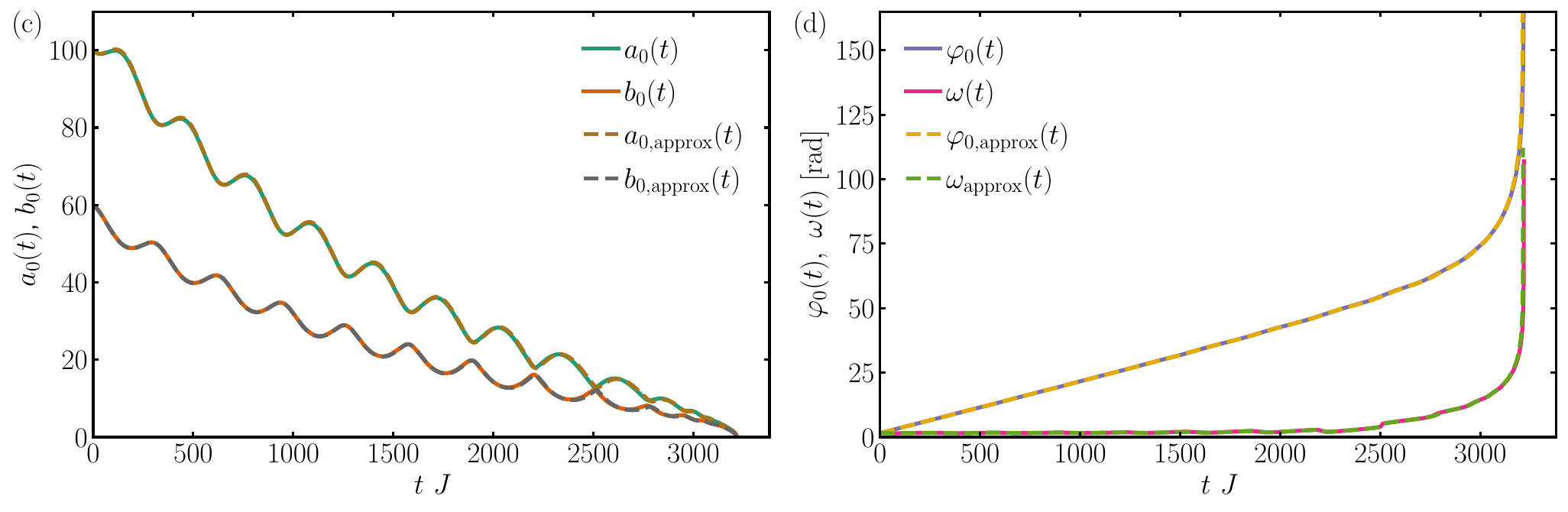}
    \caption{Time evolution of the semi-axes, helicity, and rotation angle of antiskyrmions for different initial ellipticities. Panels (a) and (b) correspond to a weakly elliptic initial state with $a_{0,0}=100$, $b_{0,0}=95$, while panels (c) and (d) show a more strongly elliptic configuration with $b_{0,0}=60$. Dashed lines in the latter two panels indicate solutions obtained from the approximate DMI contributions, which agree closely with the solution obtained from the full expressions. The semi-axes shrink exponentially at large radii and follow a square-root-like decay at small radii, superimposed with quadrupole-like oscillations. The helicity initially grows linearly and develops a logarithmic divergence as the semi-axes vanish. In contrast, the rotation angle $\omega(t)$ remains pinned in the early stages, with small oscillations highlighted in the inset of panel (b). As the ellipticity $\chi$ periodically decreases due to quadrupole oscillations, the potential $V(\omega)$ increases, leading to stronger oscillations of $\omega(t)$ until it unlocks and begins to rotate freely. After unlocking, $\omega(t)$ follows $\varphi_0(t)$ with its slope locked to half that of $\varphi_0(t)$, with small modulations due to the quadrupole-like breathing of the semi-axes. Parameters: $\tilde{D}=0.06$, $\tilde{B}=0.02$, $\alpha=0.1$, $\varphi_{0,0}=\pi/2$ and $\omega_0=\pi/2$.}
    \label{fig:ODESol_ellipse_finiteD}
\end{figure*}

This pinning is shown in Fig.\ \ref{fig:ODESol_ellipse_finiteD}, where numerical solutions of the coupled equations are depicted for antiskyrmions with an initial major axis of $a_{0,0}=100$. Panels (a) and (b) correspond to a weakly elliptic initial state with $b_{0,0}=95$, while (c) and (d) display the evolution for a more strongly elliptic configuration with $b_{0,0}=60$. The dashed lines in the latter two panels correspond to the solution using the approximated DMI contributions to the overall dynamics, matching again with high accuracy to the full expressions. This highlights that the expansion in Eq.\ \eqref{eq:expansion} describes the dynamics even for large ellipticities, supporting the validity of the estimate which Eq.\ \eqref{eq:V(omega)} is based on.

The semi-axes still shrink exponentially at large radii and cross over to a square root-like decay at small radii, superimposed with the quadrupole-like oscillations. As expected, the helicity initially grows linearly and develops a logarithmic divergence as the semi-axes vanish.

In contrast, the rotation angle remains pinned during the initial stage. The inset of Fig.\ \ref{fig:ODESol_ellipse_finiteD} (b) highlights small oscillations of $\omega(t)$ around its initial value. These oscillations, whose period follows the quadrupole-like breathing of the semi-axes, gradually increase in amplitude. With each cycle, the oscillations of $a_0(t)$ and $b_0(t)$ periodically reduce their difference, which lowers the instantaneous ellipticity $\chi$. This reduction enhances the amplitude of $V(\omega)$ and thus drives increasingly strong oscillations of $\omega(t)$. Once the torque becomes sufficiently large, $\omega(t)$ unlocks and begins to rotate freely. From this point on, it enters a dynamical regime where its slope remains locked to half that of  $\varphi_0(t)$. Even though the quadrupole oscillations of the semi-axes periodically modulate $\chi$ and thereby the amplitude of $V(\omega)$, the rotation angle does not  get pinned, but instead continues to follow $\varphi_0(t)$, with only small oscillatory modulations superimposed. These modulations are also visible in the dynamics of the initially circular antiskyrmion in Fig.\ \ref{fig:ODESol_circular_finiteD}.

\section{Comparison with numerical simulations}\label{sec:numerics}
To test the validity of the analytical approximations discussed above, we perform direct numerical simulations of the LLG equation on a grid. We consider a two-dimensional square lattice with classical magnetic moments $\bm{n}_i$ at lattice sites $i$, each represented by a three-dimensional unit vector. The LLG equation in Eq.\ \eqref{eq:LLGlattice} can be rewritten in a numerically convenient explicit form
\begin{equation}
\label{eq:LLGexplicit}
\frac{\partial \bm{n}_i}{\partial t}=- \frac{1}{1+\alpha^2} \bm{n}_i \times \bm{B}_i^{\text{eff}} - \frac{\alpha}{1+\alpha^2} \bm{n}_i \times \left(\bm{n}_i \times \bm{B}_i^{\text{eff}} \right) , 
\end{equation}
which is used throughout the simulations. Here, $\bm{B}_i^{\text{eff}}$ denotes the effective magnetic field at site $i$, given by
\begin{equation}
\bm{B}_i^{\text{eff}}= -\frac{\partial H}{\partial\bm{n}_i} =J \sum_{j \in \braket{i}} \bm{n}_j + D \sum_{j \in \braket{i}} \left( \bm{n}_j \times \bm{d}_{ij} \right) + B \bm{z},
\end{equation}
where the sum $j \in \braket{i}$ runs over the four nearest neighbors of site $i$, and $\bm{z}$ is the unit vector along the $z$-axis. We set the exchange constant $J=1$ as the energy unit and vary the DMI strength $D$, and the external Zeeman field $B$. The damping parameter is chosen to be $\alpha=0.1$. Simulations are carried out on quadratic systems with $L_x=L_y=L$. The lattice constant is set to $a=1$. The system size $L$ serves as the fixed physical length scale. 

For the initial state, an antiskyrmion is placed on the lattice using the domain wall parametrization for the radial profile in Eq.\ \eqref{eq:RadiusFit}, with the wall width set to $w=\rho_0/1.8$. This is a general value we obtained by fitting the domain wall profile to relaxed antiskyrmions that where stabilized by anisotropic DMI. For initially elliptic antiskyrmions the semi-axes and rotation angle are introduced according to Eq.\ \eqref{eq:ellipticitiy}. Thus, the system at $t=0$ contains an artificially prepared antiskyrmion subject to the specified parameters. Because of the isotropic bulk DMI, the antiskyrmion is energetically always above the ferromagnetic ground state and therefore unstable. Therefore, it shrinks and eventually collapses, releasing excess energy in the form of magnon waves. 

The time evolution of the semi-axes and rotation angle is extracted by identifying the antiskyrmion boundary as the zero contour of the out-of-plane magnetization component $n_z$, determined using a level-set procedure at each time step. The resulting contour is then fitted with the elliptic form of $\rho_0$ in Eq.\ \eqref{eq:ellipticitiy}, yielding the fitting parameters $a_0$, $b_0$, and $\omega$.

The helicity dynamics is obtained by averaging the local azimuthal angles $\varphi_{0,i}$ across the grid
\begin{equation}
\langle \varphi_0 \rangle = \frac{\sum_{i} \mathcal{W}_i \cdot \varphi_{0,i}}{\sum_{i}\mathcal{W}_i},
\end{equation}
with weights $\mathcal{W}_i=1-\left|n_i^z\right|$. This ensures that spins aligned nearly parallel or antiparallel to the $z$-axis contribute less, since their azimuthal angle is less well defined.

In the following, we present the results of these simulations to compare with the analytical predictions. Given the assumptions underlying the analytical model discussed in Sec.~\ref{sec:EllipticTriangularAnsatz}, a quantitative agreement is not expected. Instead, the comparison is intended to assess whether the analytical approach captures the qualitative trends observed in the simulations. We analyze the decay of both circular and elliptic antiskyrmions, in the absence of DMI as well as at finite DMI. Initial parameters were set to $a_{0,0}=100$, $\varphi_{0,0}=\pi/2$ and $\omega_0=\pi/2$.

\subsection{Time-dependent antiskyrmion decay at zero DMI}
\begin{figure}[!]
    \centering
    \includegraphics[width=0.47\textwidth]{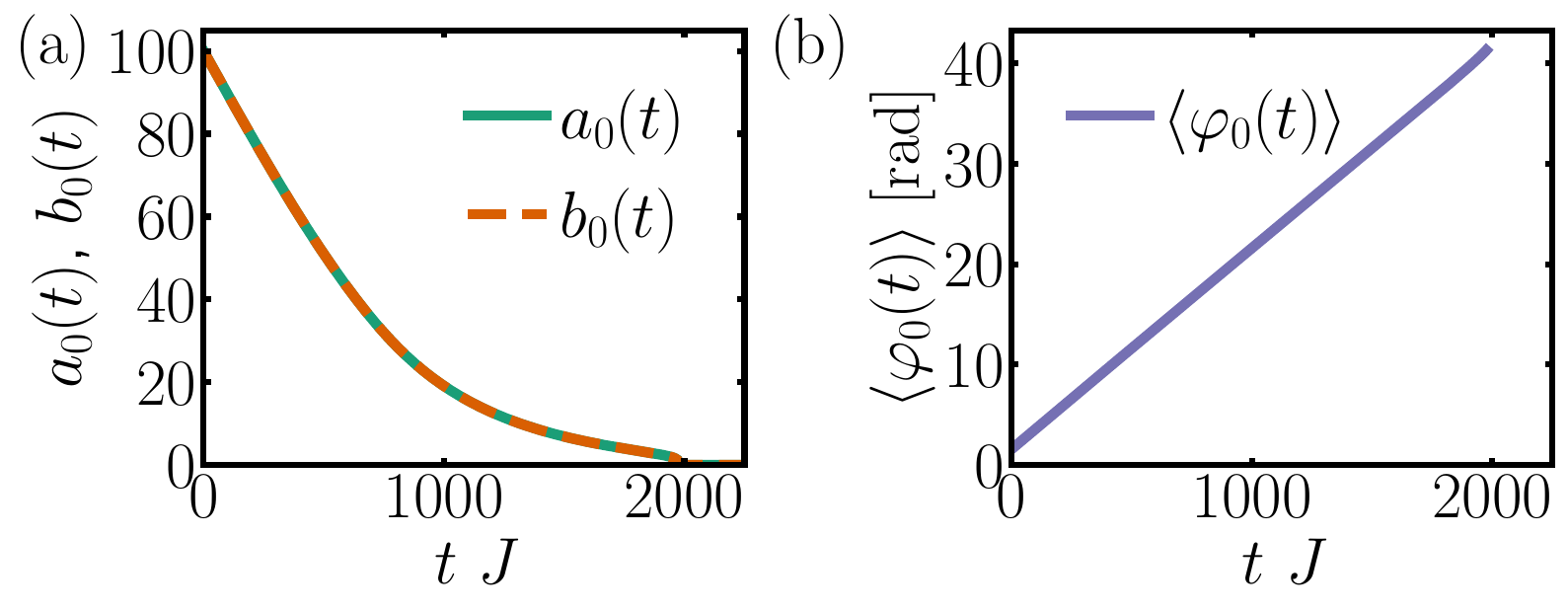}
    \includegraphics[width=0.47\textwidth]{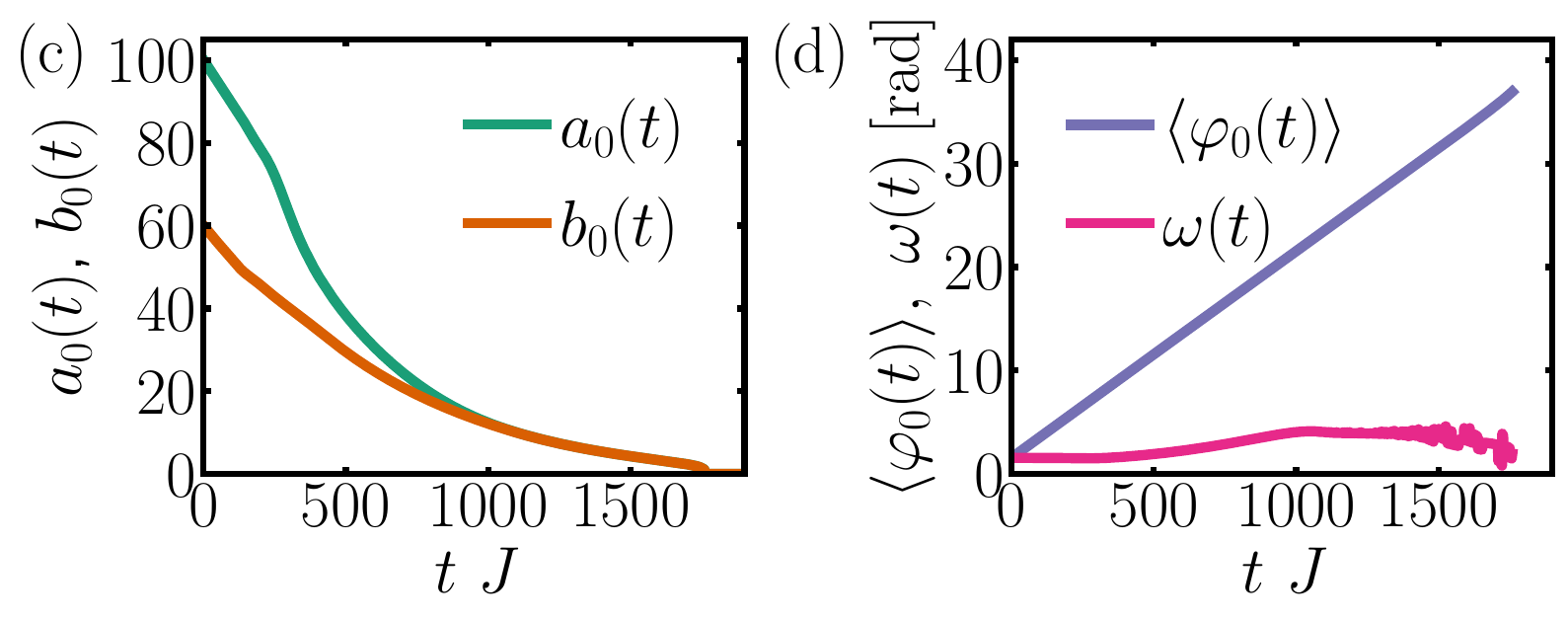}
    \caption{Time evolution of antiskyrmion decay at zero DMI. Panels (a) and (b) show the semi-axes and helicity of an initially circular antiskyrmion, while (c) and (d) display the dynamics of an initially elliptic one. In both cases the semi-axes shrink exponentially, crossing over to a square-root-like collapse at small sizes. The helicity evolves linearly in time. The initially elliptic antiskyrmion becomes circular and does not rotate to follow $\langle \varphi_0 \rangle$. Wiggles seen in $\omega(t)$, just before decay, are due to numerical difficulties to establish the directions of $a_0$ and $b_0$, when the antiskyrmion is almost circular. Parameters: $D=0$, $B=0.02$ and $\alpha=0.1$.}
    \label{fig:Simulation_D0}
\end{figure}

Figure \ref{fig:Simulation_D0} shows the dynamics of an initially circular antiskyrmion in panels (a) and (b), and of an initially elliptic one in panels (c) and (d), at vanishing DMI. In general, the system responds with a short delay to the environment, before adjusting to it.

Panel (a) demonstrates that, in this parameter regime, a circular antiskyrmion shrinks while remaining isotropic. After the initial adjustment, the semi-axes decrease exponentially, eventually crossing over to a square root-like collapse as the size becomes small. The helicity evolution in panel (b) follows a linear time dependence, driven by the external magnetic field.

The same linear dependence is observed for the helicity of the initially elliptic antiskyrmion in panel (d). In this case, the ellipticity decays, and the texture evolves into a circular antiskyrmion in (c). The semi-axes again follow an exponential law, with a square root-like collapse close to the decay. Panel (d) also shows the time evolution of $\omega(t)$. The plot confirms that at zero DMI the elliptic antiskyrmion does not rotate in-plane, in agreement with Eq.\ \eqref{eq:dotomega}. All other observations are also qualitatively consistent with the analytical predictions.

\subsection{Time-dependent antiskyrmion decay at finite DMI}
\begin{figure*}[!]
    \centering
    \includegraphics[width=\textwidth]{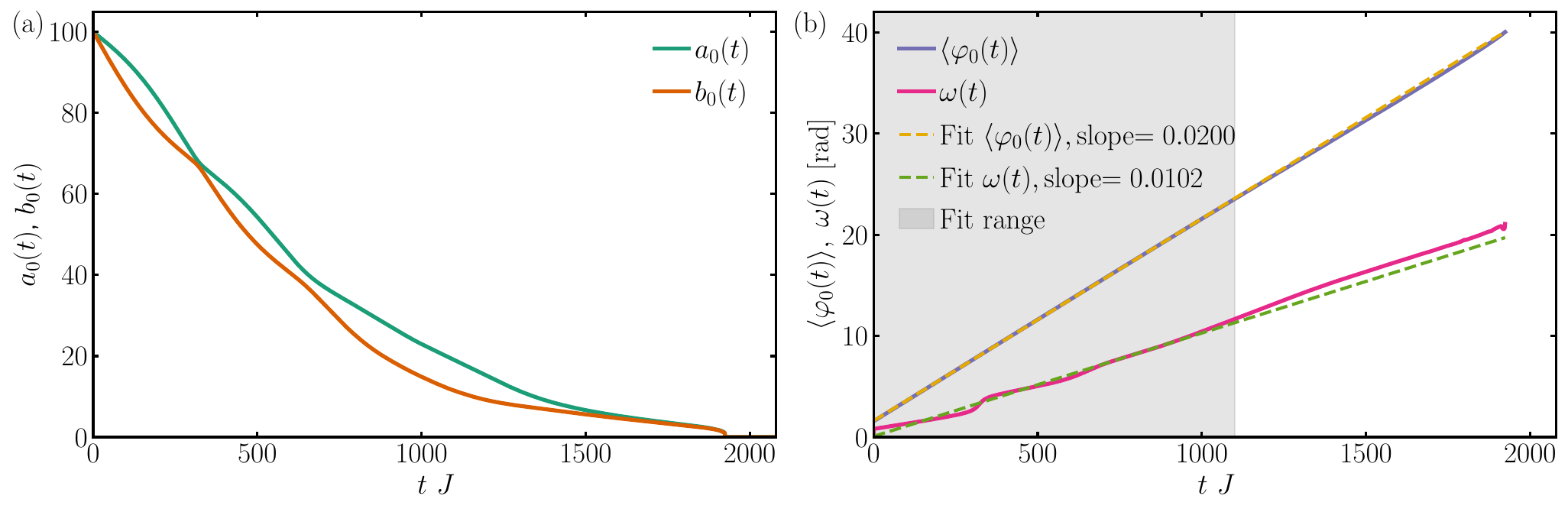}
    \includegraphics[width=\textwidth]{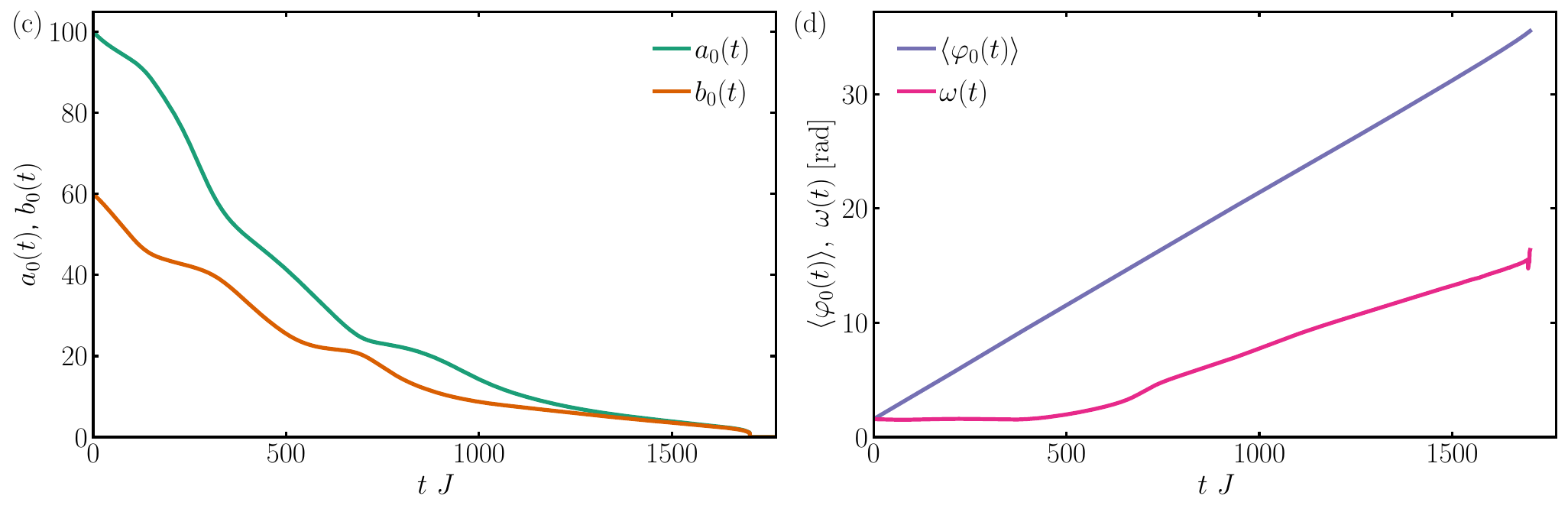}
    \caption{Time evolution of the antiskyrmion decay at finite DMI. Panels (a) and (b) show the dynamics of an initially circular antiskyrmion, which rapidly deforms into an elliptic shape with its major-axis aligning close to $\pi/4$ at short times, in agreement with Eq.\ \eqref{eq:omegaEq}. Panels (c) and (d) display the evolution of an initially elliptic antiskyrmion. In both cases, the semi-axes decay exponentially with a crossover to a square root-like collapse at small sizes, while superimposed quadrupole oscillations are visible. The linear fits shown as dashed lines in (b) show that the helicity evolves linearly in time induced by the Zeeman field strength, and the orientation angle $\omega$ follows with half the slope, as predicted. For the initially elliptic case, $\omega(t)$ is pinned until the ellipticity decreases sufficiently, such that $\omega$ unlocks and begins to rotate. Parameters: $D=0.02$, $B=0.02$, $\alpha=0.1$, $\varphi_{0,0}=\pi/2$, $\omega_0=\pi/2$.}
    \label{fig:Simulation_finiteD}
\end{figure*}

Figure \ref{fig:Simulation_finiteD} displays the dynamics of an initially circular antiskyrmion in panels (a) and (b), and of an initially elliptic one in panels (c) and (d), at finite DMI. In this regime, the decay is no longer isotropic: the circular antiskyrmion in panel (a) rapidly deforms into an elliptic shape. Its major axis aligns at short times at $\omega = 0.816 \approx \pi/4 + 0.031$, in very good agreement with the theoretical prediction from Eq.\ \eqref{eq:omegaEq}. Panel (b) additionally displays linear fits (dashed lines) to the time evolution of $\langle \varphi_0(t) \rangle$ and $\omega(t)$. As predicted, the helicity is driven by the external magnetic field, while the orientation angle $\omega$ follows $\langle \varphi_0(t) \rangle$ with half its slope.

The general time dependence of the semi-axes again exhibits an exponential decay that crosses over into a square root-like collapse at small sizes, as predicted by the theory, superimposed by quadrupole oscillations. 

The initially elliptic antiskyrmion also follows the dynamics expected from the model, including the initial pinning of $\omega(t)$, which unlocks once the ellipticity decreases sufficiently, after which $\omega$ begins to rotate with half the helicities slope.

\begin{figure*}[!]
    \centering
    \includegraphics[width=\textwidth]{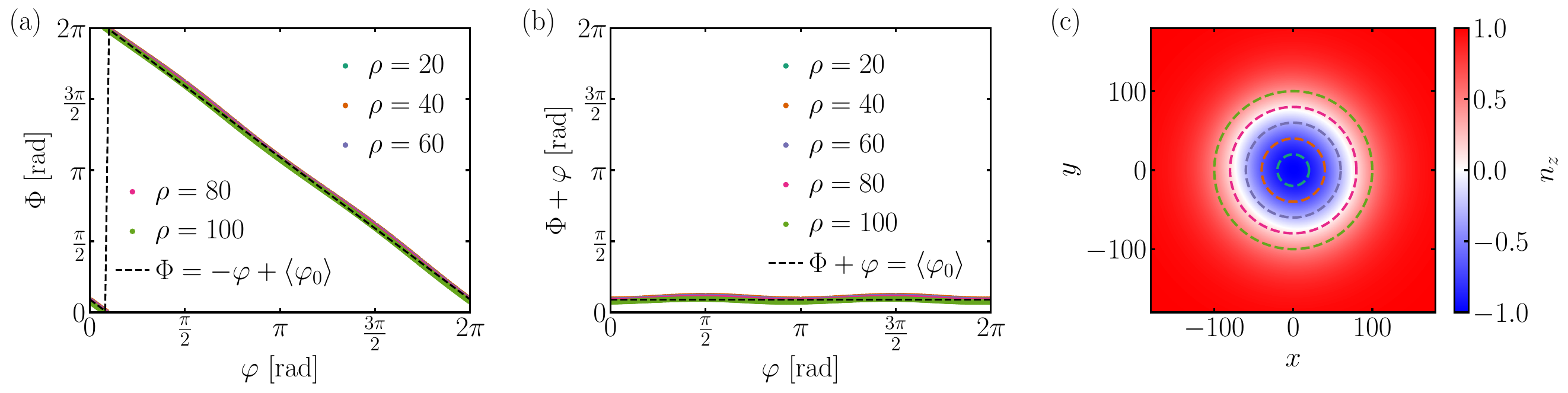}
    \includegraphics[width=\textwidth]{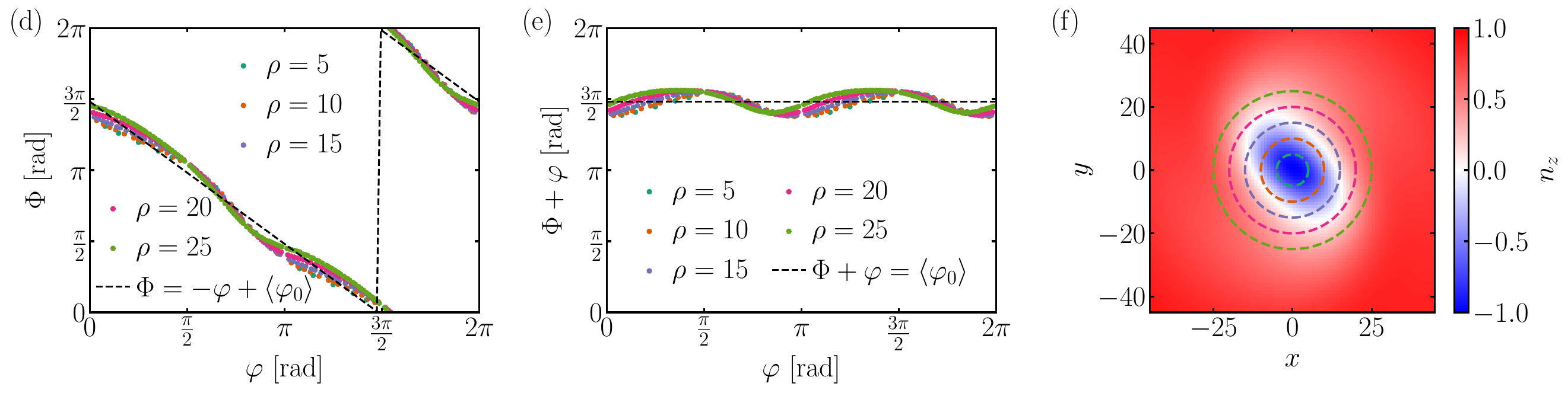}
    \caption{Angular dependence of the in-plane magnetization angle $\Phi(\varphi)$ extracted from the simulation of the antiskyrmion shrinking shown in Fig.\ \ref{fig:Simulation_finiteD} (a) and (b). Panels (a) – (c) correspond to a weakly elliptic antiskyrmion snapshot at $tJ=250$, while panels (d) – (f) show a strongly elliptic antiskyrmion at $tJ=1100$. Panels (a) and (d) display $\Phi(\varphi)$ for different values of $\rho$. Panels (b) and (e) show $\Phi(\varphi)-\varphi$ to emphasize the sinusoidal modulation around the constant averaged helicity at the respective times, indicated by the dashed black lines. Additionally, panels (c) and (f) show the corresponding $z$-component of the antiskyrmion magnetization with dashed lines indicating the values of $\rho$. In all cases, the dominant linear dependence on $\varphi$ assumed in the analytical Ansatz prevails, while a weak sinusoidal modulation reflects the elliptic deformation of the in-plane magnetization component induced by the elliptic deformation of the out-of-plane component.}
    \label{fig:Simulation_Phi}
\end{figure*}

To investigate to what extent neglecting the $\varphi$-dependent deformation of the in-plane magnetization angle $\Phi$ for elliptic antiskyrmions within the analytical approach deviates from the non-approximated dynamics, we analyze the simulation data during the shrinking process. Figure \ref{fig:Simulation_Phi} shows $\Phi(\varphi)$ for different values of $\rho$ extracted from the simulations corresponding to Fig.\ \ref{fig:Simulation_finiteD} (a) and (b). Panels (a) -- (c) correspond to a weakly elliptic antiskyrmion snapshot at $tJ=250$ and panels (d) -- (f) to a strongly elliptic one at $tJ=1100$. We observe that the dominant contribution to $\Phi(\varphi)$ remains the linear dependence on the azimuthal angle $\varphi$, as assumed in the Ansatz. This behavior is highlighted by the dashed lines, which represent the linear approximation obtained using the averaged helicity $\langle \varphi_0 \rangle$ at the respective times. This dominant linear behavior is superimposed by a weak sinusoidal modulation, which reflects the elliptic deformation of the in-plane magnetization component induced by the elliptic deformation of the out-of-plane component. This observation shows that neglecting the $\varphi$-dependent modulation of $\Phi$ within the analytical model is an approximation that disregards details of the inner structure of the antiskyrmion. For all cases studied here, however, the amplitude of this inner $\Phi$-structure never became prominent during shrinking, justifying our simplification in favor of analytical tractability and its consistency with the triangular profile assumed for $\Theta$.

Furthermore, Fig.\ \ref{fig:Simulation_Phi} (e) reveals a radial dependence of $\Phi$, as the curves corresponding to different values of $\rho$ are slightly shifted with respect to one another. While this effect is weak, it indicates that the antiskyrmion deviates from a perfectly elliptic shape during its shrinking. Such radial dependencies, generated by the full dynamics, are suppressed in our analytical model. Consequently, the Ansatz should be understood as a simplified collective-coordinate description that captures the dominant dynamical trends of antiskyrmion shrinking, rather than a quantitatively complete description of all internal deformation modes.

\section{Summary}
We developed a continuum theory for the shrinking dynamics of antiskyrmions in an environment with bulk DMI by employing a triangular elliptic approximation. Specifically, we used a triangular ansatz for the radial profile and an elliptic description for the overall shape.

The energy landscapes $E(\varphi_0,\omega)$ and $E(a_0,b_0)$ were analyzed within this framework. The first shows a $\pi$-periodic structure, independent of the DMI strength, the Zeeman field strength, and the lengths of the semi-axes, while the second reveals that antiskyrmions are always energetically unfavorable compared to the homogeneous ferromagnet. Nevertheless, elliptic antiskyrmions are energetically favored over circular ones.

From this ansatz, we derived a set of four coupled differential equations for the two semi-axes, the helicity, and the rotation angle, first at zero DMI and then at finite DMI. At zero DMI, the equations for the semi-axes decouple from those of the helicity and rotation angle, yielding constant angular velocity for the latter. The solutions show that an initially circular antiskyrmion shrinks isotropically, with an exponential law (driven by the Zeeman term) at large semi-axes and a square root-like collapse (driven by the exchange term) at small semi-axes. For an initially elliptic antiskyrmion, the system is driven towards circularity due to an antisymmetric dependence of the exchange contribution on the ratio of the semi-axes in their corresponding differential equations. Once circularity is reached, the dynamics reduces to that of the circular case. The helicity exhibits a linear time dependence at large semi-axes (Zeeman-driven) and a logarithmic divergence at small semi-axes (exchange-driven). All these predictions are consistent with micromagnetic simulations.

At finite DMI, the symmetry is broken, leading to a dynamical equation for the rotation angle and coupling between the semi-axes, helicity, and rotation angle. An initially circular antiskyrmion is driven into an elliptical shape, with its major axis aligning along half the initial helicity plus a small damping correction at short times. Again, the decay follows an exponential to square root crossover but is now superimposed with quadrupole-like oscillations of the semi-axes. At small semi-axes, the exchange term enforces circularity, reducing the dynamics to the circular case. The helicity behaves as in the zero-DMI regime, while the rotation angle follows with half the slope of the time evolution of the helicity, superimposed with step-like jumps synchronized with the oscillations of the semi-axes.

For an initially elliptic antiskyrmion at finite DMI, the rotation angle becomes pinned, oscillating around its initial value with increasing amplitude. This pinning arises from the quadrupole-like oscillations of the semi-axes, which periodically reduce their difference and thereby enhance the effective potential governing the orientation dynamics. Once the induced torque overcomes the pinning, the rotation angle resumes rotation and follows the helicity dynamics. These analytic findings are accompanied by numerical simulations confirming these behavior.

Our findings can help to understand the intrinsic dynamics of antiskyrmions used as information carriers during the writing and erasing process.

\begin{acknowledgments}
\addvspace{-0.05em}We would like to thank Martin Stier for fruitful discussions at an early stage of the research project. This work has been funded by the Deutsche For\-schungs\-gemeinschaft - Projektnummer 403505707 within the DFG Priority Program SPP 2137 ``Skyrmionics'' (Projektnummer 360506545).
\end{acknowledgments}
	
\appendix
\section{Deriving the equations of motion for zero DMI}
\label{appendixA}
As stated in the main text, insertion of the elliptic triangular ansatz into the vector field in Eq.\ \eqref{eq:vectorfield} and subsequent differentiation with respect to time yields expressions for the components of $\dot{\bm{n}}$ containing $\dot{a}_0$, $\dot{b}_0$, $\dot{\varphi}_0$, and $\dot{\omega}$, which we denote as $\dot{n}_{x,y,z}^\text{LHS}$. Another set of equations is then obtained by inserting the ansatz into the LLG equation \eqref{eq:LLG}, denoted as $\dot{n}_{x,y,z}^\text{RHS}$. By weighting the components with weighting functions and combining them, we obtain isolated equations for the time-dependent parameters. In this Appendix, we illustrate the details of this procedure. 

\subsection{Helicity}
Only the $x$- and $y$-components of the time derivative of the vector field contain terms proportional to $\dot{\varphi}_0$. Consequently, only $\dot{n}_x$ and $\dot{n}_y$ contribute to the helicity dynamics. To isolate $\dot{\varphi}_0$, the components are multiplied by the weighting functions $\sin\Phi$ and $\cos\Phi$, respectively, subtracted, and subsequently integrated over $\rho$ and $\varphi$. Remember that for antiskyrmions $\Phi=-\varphi+\varphi_0$, and we use $\eta=\pi \left( b_0^2 \cos^2\left( \varphi-\omega \right) + a_0^2 \sin^2\left(\varphi-\omega\right) \right)^2$. This yields
\begin{equation}
\begin{split}
\int_0^{2\pi} d\varphi \int_0^{2\rho_0} d\rho \rho & \left[ \dot{n}_x^{\text{LHS}} \eta \sin\Phi - \dot{n}_y^{\text{LHS}} \eta \cos\Phi \right]= \\ 
& \qquad \qquad -4 \pi a_0^2 b_0^2 \left(a_0^2 + b_0^2\right) \dot{\varphi}_0.
\end{split}
\end{equation}
Performing the same procedure with the RHS of the LLG equation yields
\begin{equation}
\begin{split}
\int_0^{2\pi} d\varphi \int_0^{2\rho_0} d\rho & \rho \left[ \dot{n}_x^{\text{RHS}} \eta \sin\Phi - \dot{n}_y^{\text{RHS}} \eta \cos\Phi \right] = \\
&\frac{a^2}{1+\alpha^2} \left\lbrace \vphantom{\frac{3}{8}}
\tilde{B} \pi \left[ -4 a_0^4 b_0^2 
- 4 a_0^2 b_0^4 \right]\right. \\
&+ J \pi^2 \left[ - 2 a_0^2 b_0^2 \pi
+ \frac{1}{4} a_0^2 b_0^2 \operatorname{Si}\!\left(2\pi\right) \right. \\
& \left. \left. + \frac{3}{8} a_0^4 \operatorname{Si}\!\left(2\pi\right)
+ \frac{3}{8} b_0^4 \operatorname{Si}\!\left(2\pi\right) \right]
\right\rbrace.
\end{split}
\end{equation}
Equating both sides yields the differential equation for the helicity
\begin{equation}
\label{eq:Adotphi0}
\begin{split}
\dot{\varphi}_0 =\frac{a^2}{1+\alpha^2}
& \left\lbrace
\tilde{B} \vphantom{\frac{3 b_0^2\,\operatorname{Si}\!\left(2\pi\right)}{32 a_0^2}} + \frac{J\pi}{a_0^2+b_0^2} \left[ 
\frac{\pi}{2}
- \frac{\operatorname{Si}\!\left(2\pi\right)}{16} \right. \right. \\
& \left. \left. - \frac{3 a_0^2\,\operatorname{Si}\!\left(2\pi\right)}{32 b_0^2}
- \frac{3 b_0^2\,\operatorname{Si}\!\left(2\pi\right)}{32 a_0^2}
\right]
\right\rbrace.
\end{split}
\end{equation}

\subsection{Semi-axes}

\subsubsection{First set of equations for semi-axes}
To derive isolated equations for $\dot{a}_0$ and $\dot{b}_0$, we use all three components of the vector field, since the $x$-, $y$-, and $z$-components are each proportional to the semi-axes. The $x$- and $y$-components are weighted by $1/\cos\xi$, while the $z$-component is weighted by $1/\sin\xi$. This choice ensures that the combination of the $x$- and $y$-contributions matches the equation obtained from the $z$-component. The fact that both approaches always yield identical results demonstrates the robustness of the method. Carrying out this procedure, the combination of the $x$- and $y$-components yields
\begin{align}
\label{eq:ALHS1}
\begin{split}
\int_0^{2\pi} d\varphi \int_0^{2\rho_0} & d\rho \rho \left[ \frac{\dot{n}_x^{\text{LHS}}}{\cos\xi} \eta \cos\Phi + \frac{\dot{n}_y^{\text{LHS}}}{\cos\xi} \eta \sin\Phi \right] = \\
& -\frac{4}{3} \pi^3 a_0 b_0 \left( b_0^3 \dot{a}_0 + a_0^3 \dot{b}_0 \right).
\end{split}
\end{align}
For the $z$-component, we straightforwardly recover the same expression
\begin{equation}
\label{eq:ALHS1_2}
\int_0^{2\pi} d\varphi \int_0^{2\rho_0} d\rho \rho \frac{\dot{n}_z^{\text{LHS}}}{\sin\xi} \eta = -\frac{4}{3} \pi^3 a_0 b_0 \left( b_0^3 \dot{a}_0 + a_0^3 \dot{b}_0 \right).
\end{equation}
For the RHS, the combination of the $x$- and $y$-components yields
\begin{equation}
\label{eq:ARHS1}
\begin{split}
\int_0^{2\pi} d\varphi \int_0^{2\rho_0} \!\! d\rho \rho & \left[ \frac{\dot{n}_x^{\text{RHS}}}{\cos\xi} \eta \cos\Phi + \frac{\dot{n}_y^{\text{RHS}}}{\cos\xi} \eta \sin\Phi \right] =  \\ 
&\frac{a^2\alpha}{1+\alpha^2}\left\lbrace
4 \tilde{B} \pi \left[a_0^4 b_0^2+a_0^2 b_0^4\right] \vphantom{\frac{3}{8}} \right. \\
&+ J\pi^2\left[
2\pi a_0^2 b_0^2
- \frac{1}{4}a_0^2 b_0^2\,\operatorname{Si}\!\left(2\pi\right) \right. \\
&\left. \left. - \frac{3}{8}a_0^4\,\operatorname{Si}\!\left(2\pi\right)
- \frac{3}{8}b_0^4\,\operatorname{Si}\!\left(2\pi\right)
\right]
\right\rbrace,
\end{split}
\end{equation}
\addvspace{2em}
\noindent and again, the $z$-component yields the identical result
\begin{equation}
\label{eq:ARHS1_2}
\begin{split}
\int_0^{2\pi} d\varphi \int_0^{2\rho_0} d\rho \rho & \frac{\dot{n}_z^{\text{RHS}}}{\sin\xi} \eta = \\ 
&\frac{a^2\alpha}{1+\alpha^2}\left\lbrace
4 \tilde{B} \pi \left[a_0^4 b_0^2+a_0^2 b_0^4\right] \vphantom{\frac{3}{8}} \right. \\
&+ J\pi^2\left[
2\pi a_0^2 b_0^2
- \frac{1}{4}a_0^2 b_0^2\,\operatorname{Si}\!\left(2\pi\right) \right. \\
&\left. \left. - \frac{3}{8}a_0^4\,\operatorname{Si}\!\left(2\pi\right)
- \frac{3}{8}b_0^4\,\operatorname{Si}\!\left(2\pi\right)
\right]
\right\rbrace.
\end{split}
\end{equation}
Thus, we obtain one relation involving both $\dot{a}_0$ and $\dot{b}_0$. In the next step, additional weighting functions are used to derive a second equation proportional to the semi-axes in order to isolate each time derivative.

\subsubsection{Second set of equations for semi-axes}
The second relation involving the two time derivatives $\dot{a}_0$ and $\dot{b}_0$ can again be obtained either from a combination of the $x$- and $y$-components or directly from the $z$-component. Therefore, the components are projected along additional weighting functions $\cos(2\Phi)/\sin(2(\varphi_0-\omega))$ and $\sin(2\Phi)/\cos(2(\varphi_0-\omega))$. From the $x$- and $y$-components we obtain the LHS
\begin{widetext}
\begin{equation}
\label{eq:ALHS2}
\begin{split}
\int_0^{2\pi} d\varphi \int_0^{2\rho_0} d\rho \rho & \left\lbrace \left[ \frac{\dot{n}_x^{\text{LHS}}}{\cos\xi} \cos\Phi + \frac{\dot{n}_y^{\text{LHS}}}{\cos\xi} \sin\Phi \right] \frac{\eta \cos(2\Phi)}{\sin(2(\varphi_0-\omega))} \right. \\
& \left. + \left[ \frac{\dot{n}_x^{\text{LHS}}}{\cos\xi} \cos\Phi  + \frac{\dot{n}_y^{\text{LHS}}}{\cos\xi} \sin\Phi \right] \frac{\eta \sin(2\Phi)}{\cos(2(\varphi_0-\omega))} \right\rbrace = \frac{4}{3}\pi^{3} a_{0} b_0\, \csc\!\left(4\left(\varphi_0-\omega\right)\right)
\left(a_0^{3}\dot{b}_0 -b_0^{3}\dot{a}_0 \right),
\end{split}
\end{equation}
while the $z$-component yields the same expression
\begin{equation}
\label{eq:ALHS2_2}
\begin{split}
&\int_0^{2\pi} d\varphi \int_0^{2\rho_0} d\rho \rho \left[ \frac{\dot{n}_z^{\text{LHS}}}{\sin\xi} \frac{\eta \cos(2\Phi)}{\sin(2(\varphi_0 - \omega))} + \frac{\dot{n}_z^{\text{LHS}}}{\sin\xi} \frac{\eta \sin(2\Phi)}{\cos(2(\varphi_0 - \omega))} \right] = \frac{4}{3}\pi^{3} a_{0} b_0\, \csc\!\left(4\left(\varphi_0-\omega\right)\right)
\left(a_0^{3}\dot{b}_0 -b_0^{3}\dot{a}_0 \right).
\end{split}
\end{equation}
The corresponding RHS reads, from the $x$- and $y$-components,
\begin{equation}
\label{eq:ARHS2}
\begin{split}
\int_0^{2\pi} d\varphi \int_0^{2\rho_0} & d\rho \rho \left\lbrace \left[ \frac{\dot{n}_x^{\text{RHS}}}{\cos\xi} \cos\Phi + \frac{\dot{n}_y^{\text{RHS}}}{\cos\xi} \sin\Phi \right] \frac{\eta \cos(2\Phi)}{\sin(2(\varphi_0-\omega))}\right. \\
& \left. + \left[ \frac{\dot{n}_x^{\text{RHS}}}{\cos\xi} \cos\Phi  + \frac{\dot{n}_y^{\text{RHS}}}{\cos\xi} \sin\Phi \right] \frac{\eta \sin(2\Phi)}{\cos(2(\varphi_0-\omega))} \right\rbrace = \frac{a^2 \alpha}{1+\alpha^2} \, \csc\!\left(4\left(\varphi_{0}-\omega\right)\right) 
\left\lbrace
4\tilde{B}\pi \left[ a_0^2 b_0^4 - a_0^4  b_0^2 \right] \vphantom{\frac{J}{2}} \right. \\
& \kern22em \left. +\frac{J}{2}\pi^2  \, \operatorname{Si}\!\left(2\pi\right)\left[a_0^4 - b_0^4\right]
\right\rbrace,
\end{split}
\end{equation}
and the $z$-component again yields the same result
\begin{equation}
\label{eq:ARHS2_2}
\begin{split}
\int_0^{2\pi} d\varphi \int_0^{2\rho_0} d\rho \rho \left[ \frac{\dot{n}_z^{\text{RHS}}}{\sin\xi} \frac{\eta \cos(2\Phi)}{\sin(2(\varphi_0 - \omega))} + \frac{\dot{n}_z^{\text{RHS}}}{\sin\xi} \frac{\eta \sin(2\Phi)}{\cos(2(\varphi_0 - \omega))} \right] =&  \frac{a^2 \alpha}{1+\alpha^2} \, \csc\!\left(4\left(\varphi_{0}-\omega\right)\right) 
\left\lbrace
4\tilde{B}\pi \left[ a_0^2 b_0^4 - a_0^4  b_0^2 \right] \vphantom{\frac{J}{2}} \right. \\
& \left. +\frac{J}{2}\pi^2  \, \operatorname{Si}\!\left(2\pi\right)\left[a_0^4 - b_0^4\right]
\right\rbrace.
\end{split}
\end{equation}
\end{widetext}
This provides a second independent relation of $\dot{a}_0$ and $\dot{b}_0$. Together with Eq.~\eqref{eq:ALHS1}--\eqref{eq:ARHS1_2}, the two time derivatives of the semi-axes can now be isolated. This is done in the following section.

\subsubsection{Isolate $\dot{a}_0$ and $\dot{b}_0$}
By taking a suitable linear combination of Eqs. \eqref{eq:ALHS1} and \eqref{eq:ALHS2}, the time derivative $\dot{a}_0$ can be isolated as
\begin{equation}
\label{eq:Aisolatedota0}
\dot{a}_0= -\frac{3}{8 \pi^3 a_0 b_0^4}\left( \text{Eq.\ } \eqref{eq:ALHS1} + \frac{\text{Eq.\ } \eqref{eq:ALHS2}}{\csc\left(4 (\varphi_0 - \omega)\right)} \right).
\end{equation}
Doing the same operations to the corresponding right hand sides Eq. \eqref{eq:ARHS1} and \eqref{eq:ARHS2} will then give us the isolated equation for $\dot{a}_0$
\begin{align}
\begin{split}
\label{eq:Adota0}
\dot{a}_0 
&= -\frac{3}{8 \pi^3 a_0 b_0^4}\left( \text{Eq.\ } \eqref{eq:ARHS1} 
   + \frac{\text{Eq.\ } \eqref{eq:ARHS2}}{\csc\!\left(4 (\varphi_0 - \omega)\right)} \right) \\
=& \frac{a^2 \alpha}{1+\alpha^2}\left\lbrace 
   -3 \tilde{B}\frac{a_0}{\pi^2} 
   + J \left[ -\frac{3 a_0}{4 b_0^2} 
   + \frac{21}{64 a_0 \pi}\,\operatorname{Si}\!\left(2\pi\right) \right. \right. \\
& \qquad \qquad \left. \left.
   -\frac{3 a_0^3}{64 b_0^4 \pi}\,\operatorname{Si}\!\left(2\pi\right)
   + \frac{3 a_0}{32 b_0^2 \pi}\,\operatorname{Si}\!\left(2\pi\right) \right]
   \right\rbrace.
\end{split}
\end{align}
The expression for $\dot{b}_0$ follows by an analogous procedure, leading to
\begin{equation}
\label{eq:Adotb0}
\begin{split}
\dot{b}_0 =&\frac{a^2 \alpha}{1+\alpha^2}\left\lbrace -3 \tilde{B}\frac{b_0}{\pi^2} 
+ J \left[ -\frac{3 b_0}{4 a_0^2} + \frac{21}{64 b_0 \pi}\,\operatorname{Si}\!\left(2\pi\right) \right. \right. \\
& \qquad \qquad \left. \left. -\frac{3 b_0^3}{64 a_0^4 \pi}\,\operatorname{Si}\!\left(2\pi\right)
+\frac{3 b_0}{32 a_0^2 \pi}\,\operatorname{Si}\!\left(2\pi\right) \right]
\right\rbrace.
\end{split}
\end{equation}

\subsection{Rotation angle}
Analogous to the procedure used for deriving the second relation involving $\dot{a}_0$ and $\dot{b}_0$, we can construct an equation that isolates $\dot{\omega}$. Here, the components are projected along the additional weighting functions $\sin(2\Phi)/\sin(2(\varphi_0-\omega))$ and $\cos(2\Phi)/\cos(2(\varphi_0-\omega))$. Combination of the $x$- and $y$-component yields
\begin{widetext}
\begin{equation}
\begin{split}
\int_0^{2\pi} d\varphi \int_0^{2\rho_0} d\rho \rho &\left\lbrace \left[ \frac{\dot{n}_x^{\text{LHS}}}{\cos\xi} \cos\Phi + \frac{\dot{n}_y^{\text{LHS}}}{\cos\xi} \sin\Phi \right] \frac{\eta \sin(2\Phi)}{\sin(2(\varphi_0-\omega))} \right. \\
&\left. - \left[ \frac{\dot{n}_x^{\text{LHS}}}{\cos\xi} \cos\Phi  + \frac{\dot{n}_y^{\text{LHS}}}{\cos\xi} \sin\Phi \right] \frac{\eta \cos(2\Phi)}{\cos(2(\varphi_0-\omega))} \right\rbrace = \frac{4}{3} \pi^3 a_0^2 b_0^2 \left(a_0^2 - b_0^2\right) \csc\left(4\left(\varphi_0 - \omega\right)\right) \dot{\omega}.
\end{split}
\end{equation}

\addvspace{1.5em}
\noindent As before, the $z$-component leads to the identical expression
\addvspace{0.5em}
\begin{equation}
\begin{split}
&\int_0^{2\pi} d\varphi \int_0^{2\rho_0} d\rho \rho \left[ \frac{\dot{n}_z^{\text{LHS}}}{\sin\xi} \frac{\eta \sin(2\Phi)}{\sin(2(\varphi_0 - \omega))} - \frac{\dot{n}_z^{\text{LHS}}}{\sin\xi} \frac{\eta \cos(2\Phi)}{\cos(2(\varphi_0 - \omega))} \right] =\frac{4}{3} \pi^3 a_0^2 b_0^2 \left(a_0^2 - b_0^2\right) \csc\left(4\left(\varphi_0 - \omega\right)\right) \dot{\omega}.
\end{split}
\end{equation}

\addvspace{1.5em}
\noindent Applying the same procedure to the RHS of the LLG equation yields for the $x$- and $y$-components
\addvspace{0.5em}
\begin{equation}
\begin{split}
\int_0^{2\pi} d\varphi \int_0^{2\rho_0} d\rho \rho &\left\lbrace \left[ \frac{\dot{n}_x^{\text{RHS}}}{\cos\xi} \cos\Phi + \frac{\dot{n}_y^{\text{RHS}}}{\cos\xi} \sin\Phi \right] \frac{\eta \sin(2\Phi)}{\sin(2(\varphi_0-\omega))} \right. \\
&\left. - \left[ \frac{\dot{n}_x^{\text{RHS}}}{\cos\xi} \cos\Phi  + \frac{\dot{n}_y^{\text{RHS}}}{\cos\xi} \sin\Phi \right] \frac{\eta \cos(2\Phi)}{\cos(2(\varphi_0-\omega))} \right\rbrace =0,
\end{split}
\end{equation}

\addvspace{1em}
\noindent and the z-component yields
\addvspace{0.5em}
\begin{equation}
\begin{split}
&\int_0^{2\pi} d\varphi \int_0^{2\rho_0} d\rho \rho \left[ \frac{\dot{n}_z^{\text{RHS}}}{\sin\xi} \frac{\eta \sin(2\Phi)}{\sin(2(\varphi_0 - \omega))} - \frac{\dot{n}_z^{\text{RHS}}}{\sin\xi} \frac{\eta \cos(2\Phi)}{\cos(2(\varphi_0 - \omega))} \right] =0.
\end{split}
\end{equation}

\addvspace{1.5em}
\noindent This shows that in the absence of DMI the rotation velocity of the ellipse vanishes, i.e. $\dot{\omega} =0$.
\end{widetext}

\section{DMI contribution to the dynamical equations}
\label{appendixB}
In this Appendix, we display the full DMI terms that contribute to the dynamical equations. Furthermore, we show how an expansion under the assumption of an antiskyrmion with small ellipticity allows for fully analytic expressions of the DMI contributions.

\subsection{Exact integral representation of the DMI contribution}
In order to derive the DMI contribution to the isolated equations for the semi-axes, the helicity and the rotation angle, we follow the steps outlined in Appendix \ref{appendixA} and apply them to the DMI terms of the LLG. Then, the DMI contributions to the system dynamics for the time dependent parameters read
\begin{widetext}
\begin{equation}
\label{eq:dota0DMI}
\begin{split}
\dot{a}_0^{\text{DMI}} =&\frac{
3 \tilde{D} a^2}{8 \sqrt{2} \, b_0^3 \, \pi^2 (1 + \alpha^2)} \int_0^{2\pi}d\varphi  \sqrt{a_0^2 + b_0^2 + (b_0^2 - a_0^2) \cos\left(2 (\varphi - \omega)\right)}
 \left[
2 (a_0^2 + b_0^2) \cos(2 \varphi - \varphi_0) \right. \\
& \left. - 2 (a_0^2 - b_0^2) \cos(\varphi_0 - 2 \omega)  + \alpha \left(a_0^2 + b_0^2 + (b_0^2 - a_0^2) \cos(2 (\varphi - \omega))\right)
\sin(2 \varphi - \varphi_0)
\right] \left(2 \cos\!\left(2 (\varphi - \omega)\right)+1 \right),
\end{split}
\end{equation}
\begin{equation}
\label{eq:dotb0DMI}
\begin{split}
\dot{b}_0^{\text{DMI}} = & - \frac{
3 \tilde{D} a^2}{8 \sqrt{2} \, a_0^3 \, \pi^2 (1 + \alpha^2)} \int_0^{2\pi}d\varphi \sqrt{a_0^2 + b_0^2 + (b_0^2 - a_0^2) \cos\left(2 (\varphi - \omega)\right)} \left[
2 (a_0^2 + b_0^2) \cos(2 \varphi - \varphi_0)\right.\\
&\left. - 2 (a_0^2 - b_0^2) \cos(\varphi_0 - 2 \omega) + \alpha \left(a_0^2 + b_0^2 + (b_0^2 - a_0^2) \cos(2 (\varphi - \omega))\right)
\sin(2 \varphi - \varphi_0)
\right] \left(2 \cos\!\left(2 (\varphi - \omega)\right) -1\right),
\end{split}
\end{equation}
\begin{equation}
\label{eq:dotphi0DMI}
\begin{split}
\dot{\varphi}_0^{\text{DMI}}= &\frac{\tilde{D} a^2}{4 \sqrt{2} \, a_0 b_0 (a_0^2 + b_0^2) (1 + \alpha^2)} \int_0^{2\pi}d\varphi  \sqrt{a_0^2 + b_0^2 + (b_0^2 - a_0^2) \cos\left(2 (\varphi - \omega)\right)} \left[
2 (a_0^2 + b_0^2) \alpha \cos(2 \varphi - \varphi_0) \right.\\
&\left.- 2 (a_0^2 - b_0^2) \alpha \cos(\varphi_0 - 2 \omega) + \left(-a_0^2 - b_0^2 + (a_0^2 - b_0^2) \cos\!\left(2 (\varphi - \omega)\right)\right)
\sin(2 \varphi - \varphi_0)
\right],
\end{split}
\end{equation}
\begin{equation}
\label{eq:dotomegaDMI}
\begin{split}
\dot{\omega}^{\text{DMI}} =& -\frac{3 \tilde{D} a^2}{2 \sqrt{2} \, a_0 b_0 (b_0^2 - a_0^2) \pi^2 (1 + \alpha^2)} \int_0^{2\pi} d\varphi  \sqrt{a_0^2 + b_0^2 + (b_0^2 - a_0^2) \cos\left(2 (\varphi - \omega)\right)} \left[
2 (a_0^2 + b_0^2) \cos(2 \varphi - \varphi_0) \right. \\
&\left. - 2 (a_0^2 - b_0^2) \cos(\varphi_0 - 2 \omega)  + \alpha \left(a_0^2 + b_0^2 + (b_0^2 - a_0^2) \cos\!\left(2 (\varphi - \omega)\right)\right)
\sin(2 \varphi - \varphi_0)
\right]
\sin\left(2 (\varphi - \omega)\right).
\end{split}
\end{equation}
\end{widetext}
Notably, the contribution of $\omega$ to the system dynamics is now nonzero. The equations of the semi-axes couple to both the helicity and the rotation angle via the DMI contribution. 

\subsection{Approximated analytical DMI contribution}
To proceed analytically, we expand the square-root under the assumption that $a_0\approx b_0$ which is true close to $t=0$ for initially circular antiskyrmions. To this end, we define $\delta=a_0^2-b_0^2$ and expand to first order, omitting terms of order $\mathcal{O}(\delta^2)$, according to 
\begin{equation}
\begin{split}
\label{eq:expansion}
&\sqrt{a_0^2 + b_0^2 - \delta \cos\left(2 (\varphi - \omega)\right)}  \\  
&\qquad\approx  \sqrt{a_0^2+b_0^2} - \frac{\delta \cos(2(\varphi-\omega))}{2\sqrt{a_0^2+b_0^2}} + \mathcal{O}(\delta^2).
\end{split}
\end{equation}
Substituting this expansion into Eqs. \eqref{eq:dota0DMI} - \eqref{eq:dotomegaDMI} makes the integrals analytically tractable. Integrating over $\varphi$ and reinserting $\delta=a_0^2-b_0^2$ yields
\begin{align}
\label{eq:dota0approx}
\begin{split}
\dot{a}_{0,\text{approx}}^{\text{DMI}} =& \frac{3 \tilde{D} a^2}{32 \sqrt{2} \pi b_0^3 \sqrt{a_0^2 + b_0^2} (1 + \alpha^2)} \times \\
& \left[ 4 (a_0^4 + 4 a_0^2 b_0^2 + 11 b_0^4) \cos(\varphi_0 - 2\omega)  \right. \\
 -& \left. (5 a_0^4 + 10 a_0^2 b_0^2 + 17 b_0^4) \alpha \sin(\varphi_0 - 2\omega) \right],
\end{split}
\end{align}
\begin{equation}
\label{eq:dotb0approx}
\begin{split}
\dot{b}_{0,\text{approx}}^{\text{DMI}} =& \frac{3 \tilde{D} a^2}{32 \sqrt{2} \pi a_0^3 \sqrt{a_0^2 + b_0^2} (1 + \alpha^2)}\times \\
&\left[ -4 (11 a_0^4 + 4 a_0^2 b_0^2 + b_0^4) \cos(\varphi_0 - 2\omega)  \right. \\
 + & \left. (17 a_0^4 + 10 a_0^2 b_0^2 + 5 b_0^4) \alpha \sin(\varphi_0 - 2\omega) \right],
\end{split}
\end{equation}
\begin{equation}
\label{eq:dotphi0approx}
\begin{split}
\dot{\varphi}_{0,\text{approx}}^{\text{DMI}} =& -\frac{\tilde{D} \pi a^2 (a_0^2 - b_0^2)}{8 \sqrt{2} a_0 b_0 \sqrt{a_0^2 + b_0^2} (1 + \alpha^2)} \times \\
& \left[ 10\alpha \cos(\varphi_0 - 2\omega) + 3 \sin(\varphi_0 - 2\omega) \right],
\end{split}
\end{equation}
\begin{equation}
\label{eq:dotomegaapprox}
\begin{split}
\dot{\omega}_\text{approx}^{\text{DMI}} =& \frac{3 \tilde{D} a^2}{16 \sqrt{2}\pi  a_0 b_0 (a_0^2 - b_0^2) \sqrt{a_0^2 + b_0^2} (1 + \alpha^2)} \times \\
& \left[ (9 a_0^4 + 14 a_0^2 b_0^2 + 9 b_0^4) \alpha \cos(\varphi_0 - 2\omega)  \right. \\
& \left. + 16 (a_0^2 + b_0^2)^2 \sin(\varphi_0 - 2\omega) \right].
\end{split}
\end{equation}


\begin{thebibliography}{2}
\bibitem{BogdanovYablonskii} A. N. Bogdanov, and D. A. Yablonskii, Thermodynamically stable ``vortices'' in magnetically ordered crystals. The mixed state of magnets, Sov. Phys. JETP, \textbf{68}, 101 (1989).

\bibitem{IvanovZhmudskii} B. A. Ivanov, V. A. Stephanovich, A. A. Zhmudskii, Magnetic vortices - The microscopic analogs of magnetic bubbles, J. Magn. Magn. Mater., \textbf{88}, 116 (1990).

\bibitem{MuhlbauerBoni} S. Mühlbauer , B. Binz, F. Jonietz, C. Pfleiderer, A. Rosch, A. Neubauer, R. Georgii,  and P. Böni, Skyrmion Lattice in a Chiral Magnet,  Science \textbf{323}, 915 (2009).

\bibitem{YuTokura} X. Z. Yu, Y. Onose, N. Kanazawa, J. H. Park, J. H. Han, Y. Matsui, N. Nagaosa, and Y. Tokura, Real-space observation of a two-dimensional skyrmion crystal, Nature \textbf{465}, 901 (2010).

\bibitem{HeinzeBlugel} S. Heinze, K. von Bergmann, M. Menzel, J. Brede, A. Kubetzka, R. Wiesendanger, G. Bihlmayer, and S. Blügel, Spontaneous atomic-scale magnetic skyrmion lattice in two dimensions,  Nature Phys \textbf{7}, 7013 (2011).

\bibitem{NagaosaTokura} N. Nagaosa and Y. Tokura, Topological properties and dynamics of magnetic skyrmions, Nat. Nanotechnol. \textbf{8}, 899 (2013).

\bibitem{IwasakiNagaosa} J. Iwasaki, M. Mochizuki, and N. Nagaosa, Universal current-velocity relation of Skyrmion motion in chiral magnets, Nature Commun. \textbf{4}, 1463 (2013).

\bibitem{ZhangHesjedal}  S. L. Zhang, W. W. Wang, D. M. Burn, H. Peng, H. Berger, A. Bauer, C. Pfleiderer, G. van der Laan, and T. Hesjedal , Manipulation of skyrmion motion by magnetic field gradients, Nature Commun. \textbf{9}, 2115 (2018).

\bibitem{QinChe} G. Qin, X. Zhang, R. Zhang, K. Pei, C. Yang, C. Xu, Y. Zhou, Y. Wu, H. Du, and R. Che, Dynamics of magnetic skyrmions driven by a temperature gradient in a chiral magnet FeGe, Phys. Rev. B \textbf{106}, 024415 (2022).

\bibitem{ChalusEskildsen} N. Chalus,  A. W. D. Leishman, R. M. Menezes, G. Longbons, U. Welp, W.-K. Kwok, J. S. White, M. Bartkowiak, R. Cubitt, Y. Liu, E. D. Bauer, M. Janoschek, M. V. Milo\ifmmode \check{s}\else \v{s}\fi{}evi\ifmmode \acute{c}\else \'{c}\fi{}, M. R. Eskildsen, Skyrmion lattice manipulation with electric currents and thermal gradients in MnSi, Phys. Rev. B \textbf{111}, 064410 (2025).

\bibitem{KiselevRossler} N. S. Kiselev, A. N. Bogdanov, R. Schäfer and U. K. Rößler, Chiral skyrmions in thin magnetic films: new objects for magnetic storage technologies?, J. Phys. D: Appl. Phys. \textbf{44}, 392001 (2011).

\bibitem{TomaselloFinocchio} R. Tomasello, E. Martinez, R. Zivieri, L. Torres, M- Carpentieri, and G. Finocchio, A strategy for the design of Skyrmion racetrack memories, Sci. Rep. \textbf{4}, 6784 (2014).

\bibitem{ZhangZhou} X. Zhang, M. Ezawa, and Y. Zhou, Magnetic skyrmion logic gates: conversion, duplication and merging of skyrmions, Sci. Rep. \textbf{5}, 9400 (2015).

\bibitem{FinocchioKlaui} G. Finocchio, F. Büttner, R. Tomasello, M. Carpentieri, and M. Kläui, Magnetic skyrmions: from fundamental to applications, J. Phys. D Appl. Phys. \textbf{49}, 423001 (2016).

\bibitem{DupeHeinze} B. Dupé, G. Bihlmayer, M. Böttcher, S. Blügel, and S. Heinze, Engineering skyrmions in transition-metal multilayers for spintronics, Nat. Commun. \textbf{7}, 11779 (2016).

\bibitem{JiangHoffmann} W. Jiang, G. Chen, K. Liu, J. Zang, S. G. E. te Velthuis, A. Hoffmann, Skyrmions in magnetic multilayers, Phys. Rep. \textbf{704}, 1 (2017).

\bibitem{Moriya} T. Moriya, Anisotropic superexchange interaction and weak ferromagnetism, Phys. Rev. \textbf{120}, 91 (1960).

\bibitem{Leonov} A. O. Leonov, T. L. Monchesky, N. Romming, A. Kubetzka, A. N. Bogdanov, and R. Wiesendanger, The properties of isolated chiral skyrmions in thin magnetic films, New J. Phys. \textbf{18}, 065003 (2016).

\bibitem{Everschor-SitteKlaui} K. Everschor-Sitte, J. Masell, R. M. Reeve, and M. Kläui, Perspective: Magnetic skyrmions --- Overview of recent progress in an active research field, J. Appl. Phys. \textbf{124}, 240901 (2018).

\bibitem{LeonovMostovoy} A. O. Leonov, and M. Mostovoy, Multiply periodic states and isolated skyrmions in an anisotropic frustrated magnet, Nat. Commun. \textbf{6}, 8275 (2015).

\bibitem{Hou} Z. P. Hou, W. J. Ren, B. Ding, G. Z. Xu, Y. Wang, B. Yang, Q. Zhang, Y. Zhang, E. K. Liu, F. Xu et al., Adv. Mater. \textbf{29}, 1701144, (2017).

\bibitem{Paul} S. Paul, S. Haldar, S. v. Malottki, S. Heinze, Role of higher-order exchange interactions for skyrmion stability, Nat. Commun. \textbf{11}, 4756 (2020).

\bibitem{Koshibae} W. Koshibae, N. Nagaosa, Theory of antiskyrmions in magnets, Nat. Commun. \textbf{7}, 10542 (2016).

\bibitem{CamosiVogel} L. Camosi, S. Rohart, O. Fruchart, S. Pizzini, M. Belmeguenai, Y. Roussigné, A. Stashkevich, S. M. Cherif, L. Ranno, M. de Santis, and J. Vogel, Anisotropic Dzyaloshinskii-Moriya interaction in ultrathin epitaxial Au/Co/W(110), Phys. Rev. B \textbf{95}, 214422 (2017).

\bibitem{Hoffmann} M. Hoffmann, B. Zimmermann, G. P. Müller, D. Schürhoff, N. S. Kiselev, C. Melcher, and S. Blügel, Antiskyrmions stabilized at interfaces by anisotropic Dzyaloshinskii-Moriya interactions,  Nat. Commun. \textbf{8}, 308 (2017).

\bibitem{Nayak} A. K. Nayak, V. Kumar, T. Ma, P. Werner, E. Pippel, R. Sahoo, F. Damay, U. K. Rößler, C. Felser, and S. S. P. Parkin, Magnetic antiskyrmions above room temperature in tetragonal Heusler materials, Nature \textbf{548}, 561 (2017).

\bibitem{CamosiRohart} L. Camosi, N. Rougemaille, O. Fruchart, J. Vogel, and S. Rohart, Micromagnetics of antiskyrmions in ultrathin films, Phys. Rev. B \textbf{97}, 134404 (2018).

\bibitem{PengTokura} L. Peng, R. Takagi, W. Koshibae, K. Shibata, K. Nakajima, T. Arima, N. Nagaosa, S. Seki, X. Yu, and Y. Tokura, Nat. Nanotechnol. \textbf{15}. 181 (2020).

\bibitem{KarubeTaguchi} K. Karube, L. Peng, J. Masell, X. Yu, F. Kagawa, Y. Tokura, and Y. Taguchi, Room-temperature antiskyrmions and sawtooth surface textures in a non-centrosymmetric magnet with S4 symmetry, Nat. Mater. \textbf{20}, 335 (2021).

\bibitem{YasinYu} F. S. Yasin, J. Masell, Y. Takahashi, T. Akashi, N. Baba, K. Karube, D. Shindo, T. Arima, Y. Taguchi, Y. Tokura, T. Tanigaki, and X. Yu, Bloch Point Quadrupole Constituting Hybrid Topological Strings Revealed with Electron Holographic Vector Field Tomography, Adv. Mater. \textbf{36}, 2311737 (2024).

\bibitem{HeShen} Z. He, Z. Li, Z. Chen, Z. Wang, J. Shen, S. Wang, C. Song, T. Zhao, J. Cai, S.-Z. Lin, Y. Zhang, and B. Shen, Experimental observation of current-driven antiskyrmion sliding in stripe domains, Nat. Mater. \textbf{23}, 1048 (2024).

\bibitem{Guang} Y. Guang, X. Zhang, Y. Liu, L. Peng, F. S. Yasin, K. Karube, D. Nakamura, N. Nagaosa, Y. Taguchi, M. Mochizuki, Y. Tokura, and X. Yu, Confined antiskyrmion motion driven by electric current excitations,  Nat Commun. \textbf{15}, 7701 (2024).

\bibitem{Stier} M. Stier, W. Häusler, T. Posske, G. Gurski, and M. Thorwart, Skyrmion-Anti-Skyrmion Pair Creation by in-Plane Currents, Phys. Rev. Lett. \textbf{118}, 267203 (2017).

\bibitem{Austrup} F. Austrup, W. Häusler, M. Lau, and M. Thorwart, Dynamics of skyrmion shrinking, Phys. Rev. B \textbf{111}, 134446 (2025).

\bibitem{McKeeverEverschor-Sitte} B. F. McKeever, D. R. Rodrigues, D. Pinna, Ar. Abanov, Jairo Sinova, and K. Everschor-Sitte, Characterizing breathing dynamics of magnetic skyrmions and antiskyrmions within the Hamiltonian formalism, Phys. Rev. B \textbf{99}, 054430 (2019).

\bibitem{LandauLifshits} L. Landau, and E. Lifshits, On the theory of the dispersion of magnetic permeability in ferromagnetic bodies, Phys. Z. Sowjet. \textbf{8}, 153 (1935).

\bibitem{Gilbert} T. L. Gilbert, A Phenomenological Theory of Damping in Ferromagnetic Materials, IEEE Trans. Magn. \textbf{40}, 3443 (2004). 

\bibitem{KronmullerFahnle} H. Kronmüller, and M. Fähnle, Micromagnetism and the Microstructure of Ferromagnetic Solids, (Cambridge University Press, Cambridge, 2003).

\bibitem{BogdanovHubert} A. Bogdanov, and A. Hubert, Thermodynamically stable magnetic vortex states in magnetic crystals, J. Magn. Magn. Mater. \textbf{138}, 255 (1994).

\bibitem{Wang} X. S. Wang, H. Y. Yuan, and X. R. Wang, A theory on skyrmion size, Commun. Phys. \textbf{1}, 31 (2018).

\bibitem{Zhuo} F. Zhuo, Z. Sun, Field-driven Domain Wall Motion in Ferromagnetic Nanowires with Bulk Dzyaloshinskii-Moriya Interaction, Sci. Rep. \textbf{6}, 25122 (2016).

\bibitem{YershovKravchuk} K. V. Yershov, A. K\'akay, and V. P. Kravchuk, Curvature-induced drift and deformation of magnetic skyrmions: Comparison of the ferromagnetic and antiferromagnetic cases, Phys. Rev. B \textbf{105}, 054425, (2022).

\bibitem{BichsKravchuk} A. Bichs, K. V. Yershov, V. P. Kravchuk, Curvature-induced skyrmion deformation, Low Temp. Phys. \textbf{51}, 550, (2025).

\bibitem{Gradshteyn} I. S. Gradshteyn, and I. M. Ryzhik, Table of Integrals, Series, and Products, 8th ed., D. Zwillinger, and V. Moll, (Academic Press, Amsterdam, 2015).

\bibitem{LauDiss} M. Lau, Dynamics of driven antiferromagnetic skyrmions, Ph.D. thesis, University of Hamburg, 2025.


\end{thebibliography}
\end{document}